\documentclass[final,5p,times,twocolumn,authoryear]{elsarticle}
\usepackage{url}
\usepackage{float}
\usepackage{hyperref}
\usepackage{multirow}
\usepackage{booktabs}
\usepackage{enumitem}
\usepackage{threeparttable}
\usepackage{siunitx}
\usepackage{lipsum}
\usepackage{lineno}
\usepackage{amsmath,amssymb,amsfonts}

\journal{Information Fusion}

\begin{document}

\begin{frontmatter}

\title{Deep learning-based astronomical multimodal data fusion: A comprehensive review}

\author[1,2,3]{Wujun Shao}
\author[1,2,3]{Dongwei Fan\corref{cor}}
\author[1,2,3]{Chenzhou Cui\corref{cor}}
\author[1,2,3]{Yunfei Xu}
\author[1,2,3]{Shirui Wei}
\author[1,2]{Xin Lyu}

\cortext[cor]{Corresponding authors\\\textit{E-mail addresses:} \url{shaowj@bao.ac.cn} (Wujun Shao), \url{fandongwei@nao.cas.cn} (Dongwei Fan), \url{ccz@bao.ac.cn} (Chenzhou Cui)}

\address[1]{National Astronomical Observatories, Chinese Academy of Sciences, Beijing 100101, China}
\address[2]{University of Chinese Academy of Sciences, Beijing 100049, China}
\address[3]{National Astronomical Data Center, Beijing 100101, China}

\begin{abstract}

With the rapid advancements in observational technologies and the widespread implementation of large-scale sky surveys, diverse electromagnetic wave data (e.g., optical and infrared) and non-electromagnetic wave data (e.g., gravitational waves) have become increasingly accessible. Astronomy has thus entered an unprecedented era of data abundance and complexity. Astronomers have long relied on unimodal data analysis to perceive the universe, but these efforts often provide only limited insights when confronted with the current massive and heterogeneous astronomical data. In this context, multimodal data fusion (MDF), as an emerging method, provides new opportunities to enhance the value of astronomical data and deepening the understanding of the universe by integrating information from different modalities. Recent progress in artificial intelligence (AI), particularly in deep learning (DL), has greatly accelerated the development of multimodal research in astronomy. Therefore, a timely review of this field is essential. This paper begins by discussing the motivation and necessity of astronomical MDF, followed by an overview of astronomical data sources and major data modalities. It then introduces representative DL models commonly used in astronomical multimodal studies, the general fusion process as well as various fusion strategies, emphasizing their characteristics, applicability, advantages, and limitations. Subsequently, the paper surveys existing astronomical multimodal studies and datasets. Finally, the discussion section synthesizes key findings, identifies potential challenges, and suggests promising directions for future research. By offering a structured overview and critical analysis, this review aims to inspire and guide researchers engaged in DL-based MDF in astronomy.
\end{abstract}


\end{frontmatter}

\section{Introduction}
\label{introduction}

In the vast and complex realm of astronomy, the acquisition and analysis of data have always been at the core of its scientific exploration. Most astronomical research, both historically and currently, relies primarily on a single type of data \citep{rizhko2025astrom3}. While these data provide valuable clues to unravelling the mysteries of the universe, they often capture only a fraction of the information needed to understand the complex nature of cosmic phenomena. The universe is a dynamic and multifaceted entity. Its behaviour is governed by multiple physical processes, which span different bands, time scales, and energy ranges. As a result, reliance on single-type data can lead to incomplete or biased interpretations, limiting our ability to reveal the mechanisms behind celestial objects and events \citep{meszaros2019multi}.

Astronomy is currently experiencing a ``data deluge'' revolution. A variety of astronomical telescopes, sky surveys, and observatories, including LAMOST, SDSS, FAST, HST, VRO and the upcoming SKA (full names provided in Table ~\ref{tab:telescopes_surveys} of ~\ref{app:abbreviation}), are collecting data with unprecedented breadth and precision \citep{huerta2019enabling}. These data span the electromagnetic spectrum (e.g., $\gamma$-ray, X-ray, ultraviolet, optical, infrared, and radio) as well as non-electromagnetic messengers such as gravitational waves, cosmic rays, and neutrinos. Originating from diverse astrophysical processes, the data are ultimately presented in multiple modalities, including images, spectra, time series, tables, and text, etc. Each modality provides unique dimensional information. Crucially, these data are not isolated; rather, they represent complementary projections of the same cosmic phenomena \citep{yu2019astronomical}. Therefore, to deepen our understanding of the universe, it is essential to break down the information barriers between different modalities and advance effective data fusion.

In fact, the concept of data fusion has long existed in astronomy. Decades ago, some forward-looking astronomers had already realized the importance of integrating observational data from different wavebands for astronomical research \citep{padovani1995connection}. Such synergistic analysis of cross-band data is often termed ``multi-band data fusion'' \citep{yu2019astronomical}. In 2017, the electromagnetic counterparts of the gravitational-wave transient GW 170817A and the high-energy neutrino event IceCube-170922A were identified for the first time \citep{meszaros2019multi}. These milestone discoveries marked the beginning of practical explorations in ``multi-messenger data fusion''. However, constrained by traditional analytical methods, most early studies relied on co-location and statistical association of preprocessed data products or derived parameters to achieve so-called ``fusion'' \citep{li2019mcatcs}. Such approaches remained at the stage of data correlation analysis and failed to model the complex nonlinear relationships inherent in the raw high-dimensional representations across modalities. The core limitation was the absence of deep feature extraction and representation learning for multimodal data, which fundamentally restricted the scientific discovery potential of these early fusion efforts.

\begin{figure*}[th!]
	\centering 
	\includegraphics[width= 1 \textwidth]{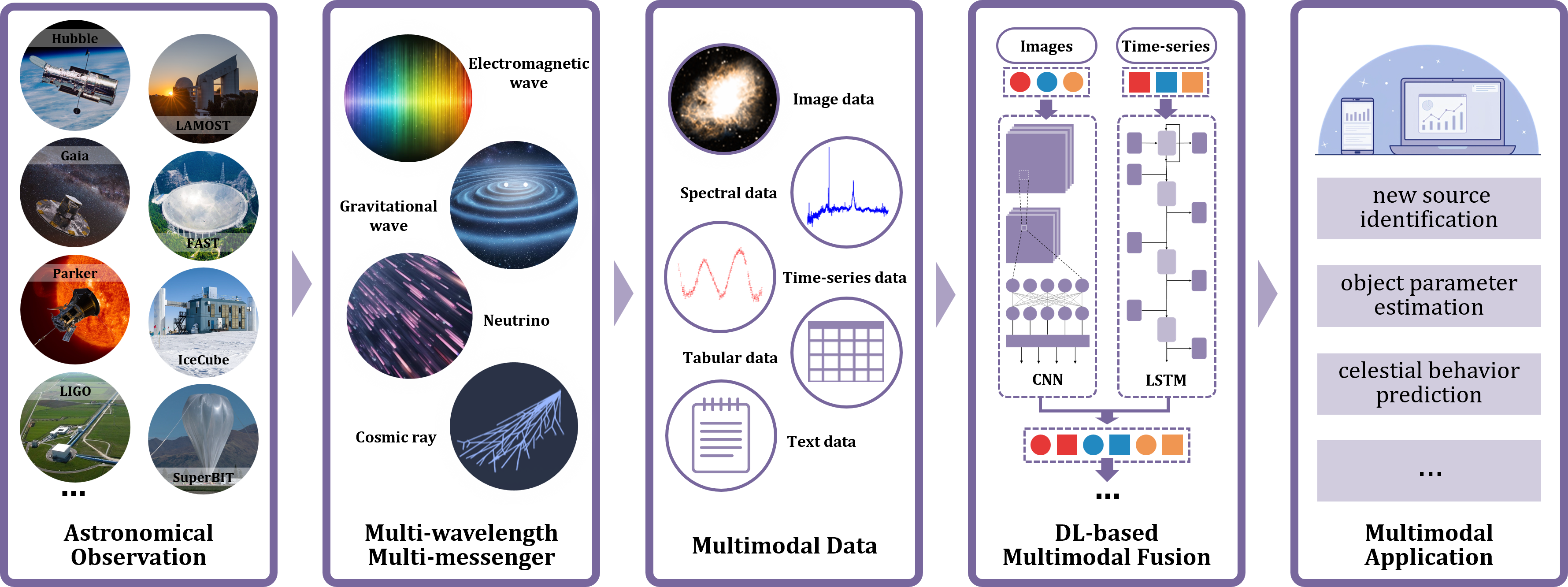}
	\caption{The basic framework of DL-based MDF in astronomy. } 
	\label{fig:framework}
\end{figure*}

This bottleneck is now being overcome by the rise of artificial intelligence (AI), particularly deep learning (DL). DL models possess powerful capabilities for automatic feature extraction and modeling complex relationships, making them an ideal tool for establishing deep connections across multimodal data. Alegre et al. \citep{alegre2024identification} develop a multimodal DL architecture combining a Convolutional Neural Network (CNN) for LOFAR radio images with an Artificial Neural Network (RNN) processing source and neighbour tabular data, achieving automatic identification of multicomponent radio sources. Leung and Bovy \citep{leung2024towards} employed a Transformer model to integrate stellar spectra, photometric data and interstellar extinction information, constructing a foundational model for stellar analysis. These studies on DL-based astronomical multimodal data fusion (MDF) can be summarized within a unified framework, as shown in Figure \ref{fig:framework}.

Despite the transformative potential of DL-based MDF to break through the limitations of existing research paradigms across numerous disciplines \citep{krones2025review}, it has yet to gain widespread adoption in astronomy. This gap arises not only from the technical challenges posed by the inherent complexity of astronomical data but more fundamentally from the path dependency and knowledge inertia established by decades of unimodal depth analysis. Against this backdrop, a comprehensive review of this emerging field is both timely and essential. Previously, while Cuoco et al. \citep{cuoco2022computational} published a review on astronomical MDF, their scope was limited to computational bottlenecks, without delving into broader concepts and methodologies. This knowledge gap offers an important entry point for this paper. 

Through a systematic search of the Astrophysics Data System (ADS) \footnote{https://ui.adsabs.harvard.edu}, Google Scholar \footnote{https://scholar.google.com} and Web of Science \footnote{https://www.webofscience.com}, we identified 58 core papers and 6 public datasets on astronomical MDF published up to October 2025. Building upon these astronomical multimodal studies, the key contributions of this review are fourfold:

\begin{itemize}[noitemsep, topsep=0pt]
\item We establish for the first time a systematic framework for DL-based astronomical MDF, unifying perspectives on data sources, modality types, fusion levels, and model development.

\item We categorize and critically analyze representative DL architectures and fusion strategies, emphasizing their mechanisms, advantages, and applicability in astronomy.

\item We provide a detailed analysis of the retrieved papers and datasets, offering the first comprehensive overview of the field.

\item We delve into current challenges such as data heterogeneity and the lack of multimodal fusion benchmarks, and propose potential future research directions to advance the field.

\end{itemize}

The remainder of this paper is structured as follows: Section \ref{sec:overview} provides an overview of the sources of astronomical data and the five primary data modalities. Section \ref{sec:method} introduces six typical DL models in astronomical MDF. Section \ref{sec:multimodal} conducts a detailed review of the general process of fusion, fusion strategies, as well as collected papers and datasets related to astronomical MDF. Section \ref{sec:discussion} presents some main findings, current challenges and future research directions. Finally, Section \ref{sec:conclusion} concludes this paper.

\section{Overview of astronomical data}\label{sec:overview}

Modern astronomy is essentially data-driven, relying on diverse observational data such as the electromagnetic spectrum. The progress of astronomical research depends not only on the amount of data collected but also on our ability to effectively interpret the data. In the context of DL-based MDF, understanding the sources and modalities of astronomical data is essential. This section will focus on two key aspects: astronomical data sources and data modalities.

\subsection{Data sources}

\begin{figure*}[th!]
	\centering 
	\includegraphics[width= 1 \textwidth]{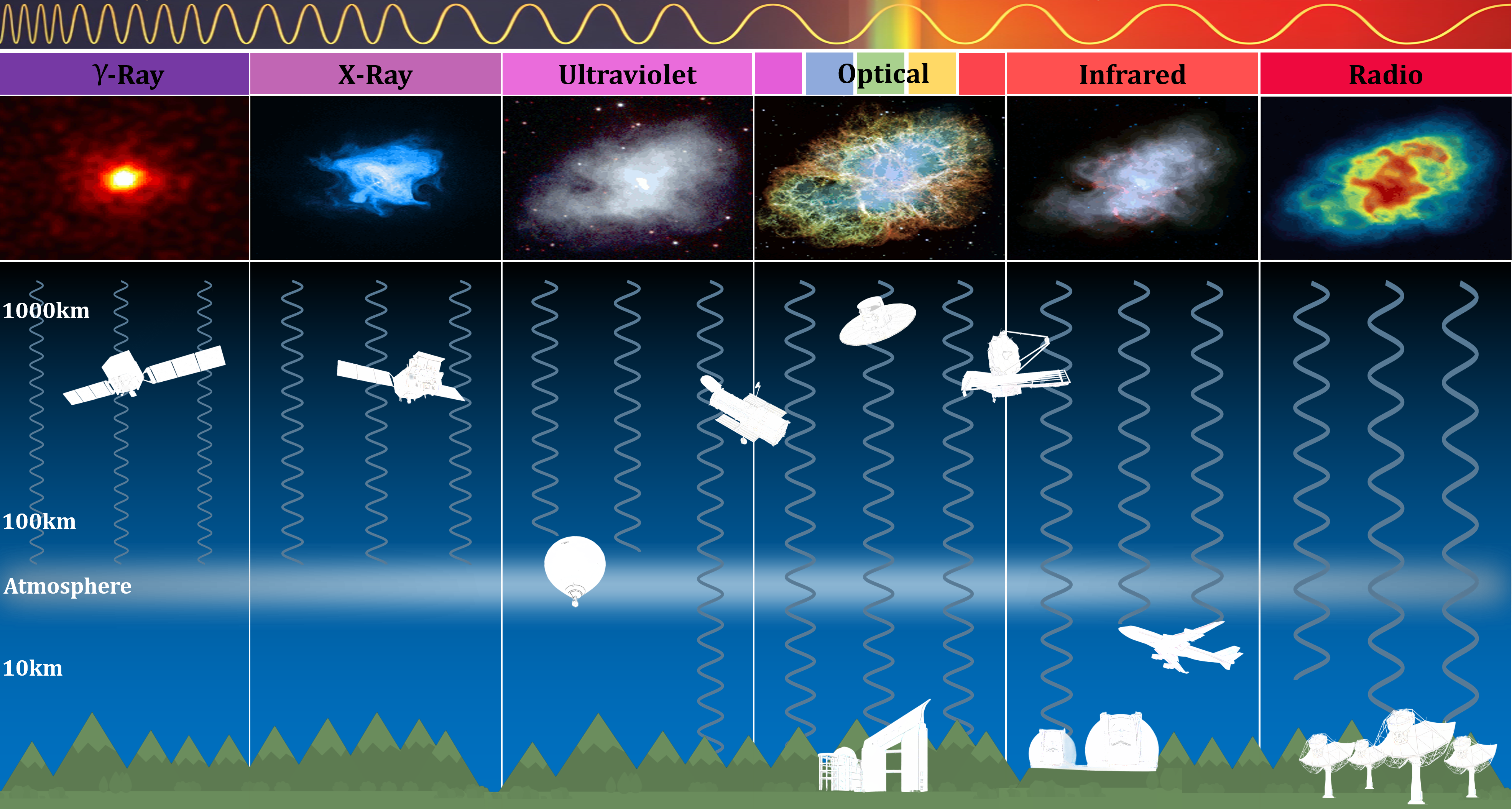}	
	\caption{
	Multi-band and multi-platform collaborative observations. The upper part: Taking multi-band observation images of the Crab Nebula as an example, it is evident that data from different wavebands can reveal distinct details, demonstrating the complementarity among them. Credit: [$\gamma$-ray: NASA/DOE/Fermi LAT/R. Buehler; X-ray: NASA/CXC/SAO/F. Seward et al.; Ultraviolet: NASA/Swift/E. Hoversten, PSU; Optical: NASA, ESA, J. Hester, and A. Loll (Arizona State University); Infrared: NASA/JPL-Caltech/R. Gehrz (University of Minnesota); Radio: NRAO/AUI and M. Bietenholz, J. M. Uson, T. J. Cornwell]. The lower part: Given that the Earth's atmosphere absorbs most of the electromagnetic radiation, only specific wavebands (primarily including the optical band, parts of the near-infrared band, and the radio band) can effectively penetrate the atmosphere and reach the ground for detection by ground-based telescopes. For other bands such as $\gamma$-ray, X-rays, and far-infrared radiation, observations must be conducted using high-altitude balloons, aircraft, or space platforms.} 
	\label{fig:multi_band}
\end{figure*}

The primary data sources for astronomical research stem from the capture and recording of various cosmic messengers. These messengers include electromagnetic radiation across all wavelengths, gravitational waves, neutrinos, and cosmic rays, each of which carries unique information about different aspects of astrophysical processes. Among them, electromagnetic wave data stand as the most crucial source for current astronomical research, owing to its long history of study, mature technological methods, and broad range of applications. The electromagnetic spectrum spans an extremely wide range, covering the full band from radio waves (long wavelength, low frequency) to $\gamma$-rays (short wavelength, high frequency). Radiation in different wavebands can reveal distinct physical processes or phenomena, thereby providing a more comprehensive perspective for astronomical observations \citep{barmby2018astronomical} (see the upper part of Figure \ref{fig:multi_band} for the multi-band images of the Crab Nebula): radio waves are well-suited for studying cold gas and pulsars; the infrared band can penetrate dust to observe stellar formation regions; the optical band provides key information on celestial morphology and stellar properties; while X-rays and $\gamma$-rays are closely associated with extreme high-energy processes such as black hole accretion and hot gas in galaxy clusters. 

To acquire these multi-band electromagnetic data, astronomers have established a collaborative observation system comprising ``ground-based, air-based, and space-based'' platforms, as shown in the lower part of Figure \ref{fig:multi_band}. Ground-based telescopes offer significant advantages in radio, optical, and near-infrared bands, achieving high-resolution observations through technologies such as large-aperture designs and adaptive optics. However, Earth's atmosphere substantially absorbs most infrared, ultraviolet, X-ray, and gamma-ray radiation. Air-based (e.g., high-altitude balloons, aircraft) and space-based platforms (e.g., HST \citep{lallo2012experience}, EP \citep{yuan2016exploring}) overcome atmospheric barriers, enabling ``unobstructed'' observations across the entire electromagnetic spectrum. Beyond the diversification of observational platforms, innovation in observational strategies represents another important dimension of data acquisition. Time-domain surveys and all-sky monitoring projects (e.g., ZTF \citep{bellm2018zwicky}, VRO \citep{ivezic2019lsst}) systematically generate vast amounts of temporal observational data by repeatedly scanning large sky regions across multiple wavebands, providing opportunities to study the dynamic evolution of the universe.

For non-electromagnetic messengers, gravitational waves (detected by facilities such as LIGO \citep{aasi2015advanced} and Virgo \citep{acernese2014advanced}) have opened entirely new windows for observing extreme cosmic events like black hole mergers; neutrinos (detected by instruments like IceCube \citep{aartsen2017icecube}) serve as direct probes for tracing nuclear reaction signatures within stellar interiors; and cosmic rays (detected by observatories such as Auger \citep{pierre2015pierre} and LHAASO \citep{cao2021ultrahigh}) provide unique messengers for exploring cosmic particle acceleration mechanisms, though their charged nature causes trajectory deflections by interstellar magnetic fields, complicating direct source tracing. Compared to electromagnetic wave data, human understanding of non-electromagnetic messengers remains in its infancy \citep{meszaros2019multi}. Nevertheless, their synergistic analysis with electromagnetic data is progressively advancing multi-messenger astronomy toward greater maturity.

\subsection{Data modalities}

All data must be presented through a modality. Modality, in a general sense, refer to the way in which something happens or is experienced \citep{baltruvsaitis2018multimodal, ngiam2011multimodal}, including images, text, video and audio. However, in specific disciplines, the scope of data types included under ``modality'' is often further extended and enriched to better align with and fulfill the needs of the field \citep{duan2024deep, luo2023artificial}. In this paper, we summarize the key modalities in astronomy, including image, spectral, time-series, tabular and text data \citep{possel2020beginner, cuoco2022computational}, as shown in Table ~\ref{tab:modalities}.

\begin{table*}[th!]
\centering
\caption{Modalities in astronomical MDF and their descriptions.}
\label{tab:modalities}
\resizebox{\textwidth}{!}{%
\begin{tabular}{@{}lllll@{}}
\toprule
  Modality &
  Primary sources &
  Characteristics &
  Core functions &
  Challenges\\ \midrule
  Image data &
  \multicolumn{1}{l}{\begin{tabular}[c]{@{}l@{}}Large-scale imaging surves\\ (e.g., SDSS, Pan-STARRS,\\ HST, VRO, JWST, Euclid)\end{tabular}} &
  \multicolumn{1}{l}{\begin{tabular}[c]{@{}l@{}}
  
  2-D pixel arrays; multi-wavelength;\\ encodes morphology \& photometry;\\ high dimensionality; large volume
  
  \end{tabular}} &
  \multicolumn{1}{l}{\begin{tabular}[c]{@{}l@{}}
  
  Provides spatial structure, morphology,\\ and photometric context; enables visual\\ interpretation and spatial analysis
  
  \end{tabular}} &
  \multicolumn{1}{l}{\begin{tabular}[c]{@{}l@{}}
  
  Resolution/calibration mismatch\\ across instruments; atmospheric\\ distortion; cross-band alignment
  
  \end{tabular}} \\ \addlinespace[1em]
  
  Spectral data &
  \multicolumn{1}{l}{\begin{tabular}[c]{@{}l@{}}Large-scale spectroscopic\\ surves (e.g., SDSS, DESI,\\ LAMOST, APOGEE)\end{tabular}} &
  \multicolumn{1}{l}{\begin{tabular}[c]{@{}l@{}}
  
  1-D intensity vs. wavelength; high\\ dimensionality; contains numerous\\ absorption lines and emission lines
  
  \end{tabular}} & \multicolumn{1}{l}{\begin{tabular}[c]{@{}l@{}}
  
  Reveals physical \& chemical properties\\ (e.g., redshift, RV, $T_{\text{eff}}$, $\log g$, elemental\\ abundances)  
  
  \end{tabular}} &
  \multicolumn{1}{l}{\begin{tabular}[c]{@{}l@{}}
  
  Requires various preprocessing;\\ high-dimensional and intricate\\ feature structure
  
  \end{tabular}} \\ \addlinespace[1em]
  Time-series data &
  \multicolumn{1}{l}{\begin{tabular}[c]{@{}l@{}}
  
  Large-scale time-domain\\ observation projects (e.g.,\\ Kepler, LIGO, IceCube)
  
  \end{tabular}} &
  \multicolumn{1}{l}{\begin{tabular}[c]{@{}l@{}}
  
  Temporal flux variations; captures\\ dynamic and transient behaviors;\\ wide dynamic range (ms → yrs)
  
  \end{tabular}} &
  \multicolumn{1}{l}{\begin{tabular}[c]{@{}l@{}}
  
  Enables study of variability, periodicity,\\ and transient events; augments static\\ multimodal data
  
  \end{tabular}} &
  \multicolumn{1}{l}{\begin{tabular}[c]{@{}l@{}}
  
  Irregular sampling, incomplete\\ coverage, heteroscedastic noise;\\ temporal alignment issues
  
  \end{tabular}}\\ \addlinespace[1em]
  Tabular data &
  \multicolumn{1}{l}{\begin{tabular}[c]{@{}l@{}}
  
   Digital sky survey catalogs\\ (e.g., Pan-STARRS, SDSS,\\ VRO, DESI)
  
  \end{tabular}} &
  \multicolumn{1}{l}{\begin{tabular}[c]{@{}l@{}}
  
  Structured rows and columns; high\\ information density; heterogeneous\\ units/scales
  
  \end{tabular}} &
  \multicolumn{1}{l}{\begin{tabular}[c]{@{}l@{}}
  
  
  Supplies metadata and labels; enables\\ cross-matching \& alignment of multi-\\modal data; links cross-survey sources
  
  \end{tabular}} &
  \multicolumn{1}{l}{\begin{tabular}[c]{@{}l@{}}
  
  Heterogeneous formats; missing\\ values; unit/scale inconsistencies;\\ needs careful integration
  
  \end{tabular}}\\ \addlinespace[1em]
  Text data&
  \multicolumn{1}{l}{\begin{tabular}[c]{@{}l@{}}
  
  Literature, observation logs,\\ alerts (e.g., SIMBAD, ADS,\\ arXiv, GCN, ATel)
  
  \end{tabular}} &
  \multicolumn{1}{l}{\begin{tabular}[c]{@{}l@{}}
  
  Unstructured or semi-structured;\\ rich in contextual and semantic\\ knowledge
  
  \end{tabular}} &
  \multicolumn{1}{l}{\begin{tabular}[c]{@{}l@{}}
  
  Provides contextual knowledge; links\\ observations to theory; supports NLP-\\driven fusion
  
  \end{tabular}} &
  \multicolumn{1}{l}{\begin{tabular}[c]{@{}l@{}}
  
  Domain-specific terminology;\\ sparse and noisy labels; needs\\ NLP \& semantic alignment
  
  \end{tabular}}\\ \bottomrule
\end{tabular}}
\end{table*}

\subsubsection{Image data}

Image data represents the most fundamental and intuitive modality in astronomy, providing direct visual evidence for exploring and understanding the universe. Among various forms of image data, imaging data occupies a central role and constitutes the largest proportion of astronomical observations, thus serving as the primary focus of this discussion. Such data are primarily acquired through telescopic observations, which record the spatial distribution of celestial radiation as two-dimensional pixel arrays encoding key observational quantities such as flux, position, and signal-to-noise ratio. From these pixel-level measurements, a range of astrophysical parameters, such as the morphology, photometric properties, and spatial correlations of celestial objects, can be systematically derived \citep{york2000sloan}. Depending on the wavelength coverage and instrumental configuration, imaging data can probe diverse physical processes: optical and infrared observations trace stellar populations and dust distributions, whereas radio and X-ray imaging reveal high-energy phenomena such as synchrotron emission and accretion dynamics. In practice, multi-band or multi-channel imaging across diverse spectral ranges naturally endows astronomical image data with intrinsic multimodal characteristics \citep{ma2017multimodal}.

Beyond its role in visual interpretation, imaging data provides a quantitative foundation for a broad spectrum of astrophysical analyses, including source detection \citep{masias2012review}, photometric measurement \citep{padmanabhan2008improved}, morphological classification \citep{spindler2021astrovader}, and astrometric calibration \citep{pier2003astrometric}. The advent of modern large-scale imaging surveys, such as SDSS, VRO, HST, JWST, Euclid, and Pan-STARRS, has further transformed the landscape of astronomical imaging. These facilities generate petabyte-scale repositories of high-resolution, multi-epoch, and multi-band images, thereby extending traditional imaging into both temporal and spectral domains. However, such data are susceptible to atmospheric disturbances, instrumental responses and background noise. Moreover, in cross-wavelength and cross-platform data integration, systematic corrections must be applied to address differences in spatial resolution, calibration methods, and noise patterns to achieve effective MDF.

\subsubsection{Spectral data}

Spectral data is another important modality in modern astronomy, revealing the physical and chemical properties of celestial objects. By dispersing incoming electromagnetic radiation into its constituent wavelengths, spectra encode rich diagnostic information about stellar atmospheres, interstellar media, and extragalactic systems. Through the analysis of absorption and emission lines \citep{possel2020beginner}, astronomers can infer parameters such as redshift, radial velocity (RV), effective temperature ($T_{\text{eff}}$), surface gravity ($\log g$), and elemental abundances \citep{rouan2015spectroscopy}, which collectively characterize the evolutionary states and kinematic properties of astronomical sources. Large-scale spectroscopic surveys, including SDSS, LAMOST, Keck, APOGEE, and DESI, have enabled the acquisition of millions of high-dimensional spectra across broad wavelength ranges and resolutions. These spectral data have fundamentally shaped our understanding of Galactic structure, stellar populations, and cosmological evolution.

From a data perspective, astronomical spectra exhibit distinctive structural and analytical characteristics compared to other modalities. Each spectrum can be viewed as a one-dimensional intensity distribution as a function of wavelength or frequency, often accompanied by metadata such as observation time, signal-to-noise ratio, and instrumental response \citep{sanchez2012califa}. The complexity of spectral features --- ranging from narrow atomic transitions to broad molecular bands --- necessitates advanced preprocessing techniques, including continuum normalization, noise reduction, and wavelength calibration \citep{bolton2012spectral}. Moreover, modern spectral analysis increasingly leverages machine learning and DL approaches for feature extraction, dimensionality reduction, and parameter estimation. In the context of MDF, spectral data serves as a crucial complement to other modalities like imaging data. For instance, imaging data provides spatial morphology and photometric context, spectral data supplies the underlying physical interpretation of those structures \citep{wei2023identification}. Integrating spectral data with other modalities thus enables a more holistic representation of astrophysical phenomena and supports the development of unified models for source classification, stellar evolution, and cosmological inference.

\subsubsection{Time-series data}

Unlike static imaging and spectral data, astronomical time-series data often records flux or other observable quantities as a function of time, revealing periodic, quasi-periodic, or stochastic variability. Such variability encodes fundamental physical processes, including stellar pulsations, rotation-induced modulation, binary interactions, transient explosive events, and accretion-driven phenomena in compact objects. Modern time-domain observation projects, such as ZTF, Kepler, LIGO, IceCube, and LHAASO, provide primary sources of time-series data. In practical applications, common types of time-series data include light curves \citep{mandel2002analytic}, pulsar timing signals in the radio band \citep{hewish1979observation}, and time-series related to $\gamma$-ray bursts \citep{margutti2011average}, gravitational wave events \citep{bailes2021gravitational}, and neutrino events \citep{hirata1987observation}. These data often exhibit differences in temporal resolution and noise level, especially on observational scales. On time scale, the variations can range from millisecond-level pulsar signals to long-term trends lasting years or even longer \citep{cabrera2024atat}; on amplitude scale, the signals may show weak periodic modulations or intense outbursts. Therefore, these characteristics of time-series data must be carefully considered during MDF.

Astronomical time-series are characterized by irregular sampling, heteroscedastic noise, and often incomplete coverage, which pose challenges for preprocessing, including detrending, outlier rejection, interpolation, and period-finding. Techniques such as periodogram analysis, autocorrelation, and Fourier transforms are traditionally employed to identify periodic or quasi-periodic signals, whereas DL models, including Recurrent Neural Networks and Transformer-based architectures, have recently demonstrated superior capability in capturing complex temporal patterns and predicting future behavior. Furthermore, time-series data are frequently integrated with other modalities in MDF frameworks to enhance classification \citep{pinciroli2023deepgravilens}, anomaly detection \citep{martinez2025augmenting}, and physical parameter estimation\citep{zhang2024maven}. This further highlights the essential role of time-series data in a comprehensive understanding of the dynamic universe.

\subsubsection{Tabular data}

Astronomical tabular data differs from image or spectral data in that it does not directly store the original observed signals, but rather records derived data that have been processed and analyzed \citep{wenger2000simbad}. It is typically organized in the form of rows and columns, where each row represents an independent sample and each column corresponds to a physical attribute or metadata. As a result, tabular data usually has a high information density, allowing it to carry rich information content within a relatively compact storage space. This characteristic makes it particularly important in processing and summarizing large-scale astronomical data, and also provides strong support for subsequent efficient management and analysis \citep{zhang2023efficient}. Moreover, it is also frequently used as input features for machine learning models to enhance the performance of classification or regression tasks \citep{zuo2024x, collister2004annz}.

The most common type of tabular data is the catalog, which includes stellar catalogs, galaxy catalogs, planetary catalogs, etc. They usually contain information such as celestial object identifiers, coordinates (e.g., right ascension, declination), magnitude, type, redshift, effective temperature, metallicity, proper motion, and morphology. Catalogs are massive in astronomy, with modern large-scale digital sky surveys such as SDSS, DESI, VRO, and Pan-STARRS being their main sources \citep{abdurro2022seventeenth, chambers2016pan}. Large-scale catalogs can provide statistical distribution patterns of celestial objects, such as color-magnitude diagrams of stars \citep{eyer2019gaia} and redshift-luminosity relationships of galaxies \citep{norberg20022df}, thereby revealing the evolutionary trends of celestial populations. Furthermore, the structured nature of catalogs enables efficient querying, filtering, and cross-matching across different surveys through parameters such as coordinates \citep{dongwei2019research}. This enables the effective association and integration of large-scale multi-source heterogeneous data, such as images, spectra, and time-series data. Thus, catalogs have also become a critical basis for aligning multimodal data \citep{angeloudi2025multimodal}. In addition to common catalog entries (e.g., photometric measurements \citep{gai2024simultaneous} and astrometric parameters \citep{rizhko2025astrom3}), other tabular data like observational metadata tables (e.g., observation time, instrument parameters, data processing versions) \citep{accomazzi2016aggregation} are also indispensable structured data recording carriers in astronomy.

\subsubsection{Text data}

Text data in astronomy encompass a broad range of non-numerical, unstructured, or semi-structured information, including scientific literature \citep{kurtz2000nasa}, observation logs \citep{rutledge1998astronomer}, mission reports, academic forum discussions \citep{chen2023radio}, archival catalogs with descriptive metadata \citep{wenger2000simbad}, and real-time alerts \citep{barthelmy1998grb}. These sources provide rich contextual knowledge that complements observational data, offering insights into observational conditions, historical discoveries, and theoretical interpretations. With the rapid growth of digital archives such as SIMBAD \footnote{https://simbad.u-strasbg.fr/simbad}, ADS, arXiv \footnote{https://arxiv.org}, GCN \footnote{https://gcn.nasa.gov}, and ATel \footnote{https://astronomerstelegram.org}, astronomical research increasingly relies on automated text mining, natural language processing (NLP), and semantic analysis techniques to extract relevant information from vast corpora. Text data can encode provenance, methodology, cross-identifications, and interpretive narratives, which are essential for connecting observational datasets to astrophysical knowledge and for supporting reproducible science.

Astronomical text data presents unique challenges due to strong contextual semantic coherence, domain-specific terminology, and implicit relationships between entities. To address these characteristics, researchers employ techniques such as named entity recognition, document embedding, relation extraction, and topic modeling to structurally organize and quantitatively analyze textual information. In MDF research, text modality serves as a critical bridge between observational measurements and theoretical understanding. Therefore, when combined with image, spectral and time-series data, textual information can enrich feature representations, guide model interpretation, and support knowledge-driven inference \citep{martinez2025augmenting}. This fusion enables more informed classification, discovery, and hypothesis testing in astrophysical research, highlighting the complementary role of human-curated knowledge in enhancing data-driven analyses \citep{ciucua2023harnessing, tanoglidis2024first, li2025astronomical}.

\section{DL models in astronomical MDF} \label{sec:method}

Similar to traditional unimodal data research, astronomers rarely construct DL models from scratch when conducting MDF research. Instead, they adopt well-established benchmark models as references and adapt them for specific tasks. This research paradigm offers two key advantages: on one hand, it significantly reduces computational resource consumption and time costs associated with redundant development; on the other hand, performance comparisons with benchmark models allow for objective validation of improvements in new models. This section will review the most representative foundational DL models in astronomical MDF, and a summary is provided in Table ~\ref{tab:models}. The aim is to assist researchers in quickly understanding and selecting the most suitable models for their specific tasks.

\begin{table*}[th!]
\centering
\caption{Summary of DL models in astronomical MDF.}
\label{tab:models}
\resizebox{\textwidth}{!}{%
\begin{tabular}{@{}lllll@{}}
\toprule
  Model &
  Applicable modalities &
  Typical applications &
  Main limitations &
  Common variants\\ \midrule
  ANNs &
  \multicolumn{1}{l}{\begin{tabular}[c]{@{}l@{}}
  
  tabular data\\
  spectral data

  \end{tabular}} &
  \multicolumn{1}{l}{\begin{tabular}[c]{@{}l@{}}
  
  General-purpose nonlinear mapping\\ for multimodal data; often used as\\ fusion layers or decision integrators
  
  \end{tabular}} &
  \multicolumn{1}{l}{\begin{tabular}[c]{@{}l@{}}
  
  Limited ability to capture spatial and\\temporal or cross-modal correlations;\\ rely on manual hyperparameter tuning
  
  \end{tabular}} &
  \multicolumn{1}{l}{\begin{tabular}[c]{@{}l@{}}
  
  Most DL models\\ (e.g. MLP).
  
  \end{tabular}}\\
  \addlinespace[0.7em]
  CNNs & 
  \multicolumn{1}{l}{\begin{tabular}[c]{@{}l@{}}
  
  image data\\spectral data\\time-series data
  
  \end{tabular}} &
  \multicolumn{1}{l}{\begin{tabular}[c]{@{}l@{}}
  
  As backbone networks for feature\\ extraction \& representation modeli-\\ng of grid-like data such as images
  
  \end{tabular}} &
  \multicolumn{1}{l}{\begin{tabular}[c]{@{}l@{}}
  
  Limited receptive field; overfitting on\\ small datasets; weak in modeling long-\\range dependencies 
  
  \end{tabular}} &
  \multicolumn{1}{l}{\begin{tabular}[c]{@{}l@{}}
  
  LeNet-5, ResNet, \\ InceptionNet, etc.
  
  \end{tabular}} \\ \addlinespace[0.7em]
  AEs &
  \multicolumn{1}{l}{\begin{tabular}[c]{@{}l@{}}
  
  image data\\ spectral data\\ time-series data
  
  \end{tabular}} &
  \multicolumn{1}{l}{\begin{tabular}[c]{@{}l@{}}
  
  Dimensionality reduction; unsuperv-\\ised feature extraction; shared latent\\ space learning; generative modeling
  
  \end{tabular}} &
  \multicolumn{1}{l}{\begin{tabular}[c]{@{}l@{}}
  
  Sensitive to latent dimension size; risk\\ of learning an incomplete or trivial\\ representation
  
  \end{tabular}} &
  \multicolumn{1}{l}{\begin{tabular}[c]{@{}l@{}}
  
  VAE, Masked AE,\\ Convolutional AE,\\
  Nouveau-VAE, etc.
  
  \end{tabular}}\\ \addlinespace[0.7em]
  GANs &
  \multicolumn{1}{l}{\begin{tabular}[c]{@{}l@{}}
  
  image data\\ spectral data\\ time-series data
  
  \end{tabular}} &
  \multicolumn{1}{l}{\begin{tabular}[c]{@{}l@{}}
  
  Data generation and augmentation;\\ synthesizing one modality from an-\\other (cross-modal synthesis)
  
  \end{tabular}} &
  \multicolumn{1}{l}{\begin{tabular}[c]{@{}l@{}}
  
  Training can be unstable and mode\\ collapse may occur;
  generated data\\ may lack physical plausibility
  
  \end{tabular}} &
  \multicolumn{1}{l}{\begin{tabular}[c]{@{}l@{}}
  
  Conditional GAN,\\ BigGAN, etc.
  
  \end{tabular}}\\ \addlinespace[0.7em]
  RNNs &
  \multicolumn{1}{l}{\begin{tabular}[c]{@{}l@{}}
  
  time-series data
  
  \end{tabular}} &
  \multicolumn{1}{l}{\begin{tabular}[c]{@{}l@{}}
  
  Modeling temporal dependencies;\\ As temporal feature encoders for\\ time-series data (e.g., light curves)
  
  \end{tabular}} &
  \multicolumn{1}{l}{\begin{tabular}[c]{@{}l@{}}
  
  Inefficient with sequential processing;\\ struggle with long-range dependencies\\ due to vanishing/exploding gradients
  
  \end{tabular}} &
  \multicolumn{1}{l}{\begin{tabular}[c]{@{}l@{}}
  
  LSTM, BiLSTM,\\ GRU, etc.
  
  \end{tabular}}\\ \addlinespace[0.7em]
  Transformers &
  \multicolumn{1}{l}{\begin{tabular}[c]{@{}l@{}}
  
  almost all\\ modalities
  
  \end{tabular}} &
  \multicolumn{1}{l}{\begin{tabular}[c]{@{}l@{}}
  
  Modeling global dependencies\\ enables excelling at cross-modal\\ alignment and joint representation\\ learning; core data fusion model
  
  \end{tabular}} &
  \multicolumn{1}{l}{\begin{tabular}[c]{@{}l@{}}
  
  Computationally intensive for long se-\\quences/high-resolution data; require\\ large-scale data for pre-training
  
  \end{tabular}} &
  \multicolumn{1}{l}{\begin{tabular}[c]{@{}l@{}}
  
  BERT, CLIP, GPT,\\ ViT, etc.
  
  \end{tabular}}\\ \bottomrule
\end{tabular}}
\end{table*}

\subsection{ANN models}

Artificial neural networks (ANNs) mark an early milestone in AI and serve as the conceptual foundation for most modern DL architectures. In astronomy, ANNs have been applied to various tasks, such as pulsar candidate selection \citep{eatough2010selection} and gravitational wave detection \citep{krastev2020real}, thereby demonstrating their ability to model nonlinear relationships in astronomical data. In current MDF research, Hong et al. \citep{hong2023photoredshift} employed an ANN model to map photometric data into spectral feature space, thereby enhancing the feature representation for the subsequent redshift prediction task; Ethiraj \& Bolla \citep{ethiraj2022classification} utilize an ANN to process tabular data within a multimodal architecture and extract its high-level features. However, classical ANN structures are relatively flat and rely on manual hyperparameter tuning, limiting their ability to capture complex spatial–temporal or cross-modal correlations. Therefore, directly employing ANNs as standalone models for specific tasks is currently uncommon. Instead, a more prevalent approach involves integrating their core components, particularly fully connected layers (FCs) assembled in the form of multilayer perceptrons (MLPs), as fusion layers for heterogeneous modality outputs or decision-level ensemble integrators \citep{wei2025photometric, hosseinzadeh2025end}.

\subsection{CNN models}

Convolutional Neural Networks (CNNs) \cite{lecun1998gradient}, first exemplified by LeNet-5 \citep{goodfellow2016deep}, have become a cornerstone of DL, particularly suited for analyzing grid-structured data such as astronomical images. Through their hierarchical feature-extraction mechanism, CNNs can effectively identify morphological and structural patterns in images, yielding excellent performance on astronomical tasks like morphology classification \citep{cavanagh2021morphological}, transient detection \citep{acero2023s}, and source segmentation \citep{xu2024surveying}. In multimodal learning scenarios, CNNs often serve as backbone networks for feature extraction and representation modeling of modalities such as images and spectra. Pinciroli Vago \& Fraternali \citep{pinciroli2023deepgravilens} utilize CNNs to extract localized spatial features of gravitational lensing (e.g., arcs and rings), while Rehemtulla et al. \citep{rehemtulla2024zwicky} utilized CNNs to capture bright-transient signatures (e.g., supernova eruptions) within astronomical cutouts. However, the receptive fields and gradient flows of standard CNNs are typically limited, making it difficult to mine global or long-range astronomical patterns. Subsequent advanced architectures (e.g., ResNet \citep{he2016deep}, InceptionNet \citep{szegedy2015going}) further enhanced CNN expressiveness and training stability by improving gradient propagation and enabling multiscale feature learning. Nevertheless, CNNs still face many challenges when handling multimodal data, especially in bridging spatial information with non-image modalities and mitigating overfitting under limited data. Consequently, recent studies increasingly combine CNNs with other DL architectures (e.g., Transformers) to explore more robust astronomical multimodal representation systems \citep{li2024solar}.

\subsection{AE models}

Autoencoders (AEs) are unsupervised neural networks that learn compact feature representations by reconstructing input data through an encoder-decoder framework. Their exceptional ability to capture both linear and nonlinear relationships makes them widely applicable for dimensionality reduction \citep{wang2016auto}, \citep{wang2016auto} and feature extraction \citep{pang2022masked} across various astronomical data modalities, including spectral, image, and time-series data. Furthermore, in astronomical MDF tasks, AEs facilitate the integration of heterogeneous modalities by projecting them into a shared latent space, thereby uncovering cross-modal shared features and enhancing analytical robustness. Variants like variational AEs (VAEs) further introduce probabilistic modeling to ensure latent space continuity and regularization, significantly improving generation quality and interpolation stability \citep{kingma2013auto}. This is particularly important for achieving high-quality cross-modal fusion \citep{walsh2024foundation}. Nevertheless, designing effective AEs for astronomical MDF usually requires careful tuning of latent dimensions and network capacity to balance reconstruction fidelity with representation compactness.

\subsection{GAN models}

Generative Adversarial Networks (GANs) demonstrate powerful data generation capabilities through the adversarial training dynamics between generator and discriminator networks. They are suited for astronomical data augmentation and cross-modal synthesis, especially when high-fidelity reconstruction or simulation of missing modalities is required. Geyer et al. \citep{geyer2023deep} employed GANs to generate radio interferometric images of galaxies, effectively expanding training datasets; García-Jara et al. \citep{garcia2022improving} synthesized rare light curve samples to address data imbalance; while Wang et al. \citep{wang2023multimodal} proposed a GAN-based framework to generate corresponding simulated spectral features from images, enabling knowledge transfer across modalities. Compared to standard AEs, GANs typically model more complex data distributions and thus produce visually more realistic outputs \citep{zhao2023can}. This offers significant advantages for bridging the modality gaps in MDF. However, training instability and sensitivity to high-dimensional astronomical data may limit their scalability and physical interpretability. Hybrid architectures combining GANs with diffusion-based models are emerging as promising directions to enhance generative fidelity and modality alignment in MDF \citep{song2024improving, campagne2025galaxy}.

\subsection{RNN models}

Recurrent Neural Networks (RNNs) can effectively capture the dependencies within sequential data through their internal recurrent connection structures. This characteristic grants them unique advantages in astronomical time-series analysis, such as predicting periodic phenomena \citep{kang2023periodic}, detecting stellar flare events \citep{vida2021finding}, or classifying transient celestial objects \citep{burhanudin2021light}. When extended to MDF tasks, similar to the aforementioned DL, RNNs primarily serve as feature 
encoders for time-series data \citep{yang2017deep}. However, the features of astronomical time-series data often require more finer-grained long-term contextual information to support cross-modal semantic alignment. Due to their relatively simple recurrent structures, fundamental RNNs often fall short in addressing long-range dependencies. Therefore, advanced RNN-based variants are increasingly applied in MDF tasks. Sun et al. \citep{sun2022accurate} used a Bidirectional Long Short-Term Memory (BiLSTM) to encode long-term dependencies in multivariate solar-wind sequences, while Pinciroli Vago \& Fraternali \citep{pinciroli2023deepgravilens} adopt a Gated Recurrent Unit (GRU) that simultaneously tracks short-term flares and long-term trends in light curves.

\subsection{Transformer models}

The emergence of the Transformer architecture \citep{vaswani2017attention} has revolutionized the processing of complex and heterogeneous data. Its core self-attention mechanism captures global dependencies in a parallelized manner, endowing the model with superior capabilities for modeling long-range relationships and greater scalability compared to previous DL models. Based on this, Transformers have achieved state-of-the-art performance across numerous astronomical unimodal tasks \citep{donoso2023astromer, zuo2025falco}, which has also rapidly made them a focal point in MDF \citep{cabrera2024atat}. Transformer's powerful ability in joint multimodal representation learning makes it frequently deployed in the data fusion stage \citep{wei2023identification, gao2023deep}. For instance, \citet{wei2023identification} leverage cross-attention to adaptively weight image features based on spectral information, effectively guiding the model to focus on image regions relevant to the spectra. Self-attention then further integrates these cross-modal features for global context modeling. Vision Transformers (ViTs) \citep{dosovitskiy2020image} have further enhanced their performance in astronomical image modality analysis, outperforming traditional CNN-based models in most cases \citep{parker2024astroclip}. To further enhance data fusion efficiency, multimodal Transformers such as Contrastive Language-Image Pre-training (CLIP) \citep{radford2021learning} have been applied to astronomical cross-modal retrieval tasks \citep{mishra2024paperclip, rizhko2025astrom3}. Originally developed for natural language or vision tasks, these large-scale Transformer-based pre-trained models can be fine-tuned for astronomy with limited labeled samples, mitigating challenges posed by modality heterogeneity and observational biases. This opens up new technical pathways for astronomical MDF.

\section{DL-based astronomical MDF}\label{sec:multimodal} \label{sec:multimodal_fusion}

Unimodal data inherently limit the original multidimensional expression of an object or event. Although the related analysis techniques are still evolving and may continue to help astronomers improve task performance to some extent, the ``one-sided'' nature of the data itself can be a major bottleneck. In recent years, with the rapid development of DL models and the recognition that fusing different modalities can improve the accuracy of results or decisions, DL-based MDF has become a rapidly emerging research direction in many fields. Astronomy is no exception. To provide readers with a comprehensive understanding of MDF in astronomy, this section will outline the fusion strategies (Subsection \ref{subsec:fusion_strategy}), the general process of multimodal model development (Subsection \ref{subsec:development}), multimodal related studies (Subsection \ref{subsec:study}) and datasets (Subsection \ref{subsec:dataset}) in astronomy. 
Notably, the development of multimodal models is largely influenced by fusion strategies; hence, we will first introduce multimodal fusion strategies.

\subsection{Fusion strategies} \label{subsec:fusion_strategy}

Existing studies generally categorize MDF into four primary levels: data-level fusion, feature-level fusion, decision-level fusion, and hybrid fusion. Each fusion strategy focuses on integrating information at a different stage, thereby shaping model design and application practices for effective use of multimodal data. In this paper, we adopt the same four-category framework for organization. Below, we will systematically elaborate on the specific implementation methods and representative applications of each fusion strategy, with Table ~\ref{tab:fusion_strategies} summarizing their applicable scenarios and limitations for clear reference.

\begin{table*}[ht!]
\centering
\caption{Comparison of four fusion strategies in astronomical MDF.}
\label{tab:fusion_strategies}
\footnotesize
\resizebox{\textwidth}{!}{
\begin{tabular}{@{}llll@{}}
\toprule
Fusion strategy & Characteristics & Applicable scenarios & Main limitations\\
\midrule

Data-level &
\multicolumn{1}{l}{\begin{tabular}[c]{@{}l@{}}

$\bullet$ Directly integrates raw data from\\
different modalities, preserving the\\ most complete information\\
$\bullet$ Fusion occurs before model input,\\ enabling a unified feature space

\end{tabular}} & 

\multicolumn{1}{l}{\begin{tabular}[c]{@{}l@{}}

$\bullet$ Highly correlated modalities (e.g., multi-\\band images)\\
$\bullet$ Well-aligned data with similar structures

\end{tabular}} &
\multicolumn{1}{l}{\begin{tabular}[c]{@{}l@{}}

$\bullet$ Struggles with semantic and statistical heterogeneity\\
$\bullet$ Assumes equal modality importance and independence\\
$\bullet$ Prone to noise and the ``curse of dimensionality''

\end{tabular}}\\
\addlinespace[1em]

Feature-level &
\multicolumn{1}{l}{\begin{tabular}[c]{@{}l@{}}

$\bullet$ Extracts features from each\\ modality before fusion, promoting\\ deep cross-modal interactions\\
$\bullet$ Balances modality-specific\\ feature extraction with cross-modal\\ relationship modeling

\end{tabular}} &
\multicolumn{1}{l}{\begin{tabular}[c]{@{}l@{}}

$\bullet$ Structurally heterogeneous but semantically\\ consistent multimodal data\\
$\bullet$ Tasks requiring deep feature extraction and\\ modeling

\end{tabular}} &
\multicolumn{1}{l}{\begin{tabular}[c]{@{}l@{}}

$\bullet$ Relies heavily on the quality of feature extraction\\
$\bullet$ Assumptions of semantic alignment between modalities\\ may not hold\\
$\bullet$ Struggles with asynchronous observations or significant\\ spatio-temporal resolution differences

\end{tabular}}\\
\addlinespace[1em]

Decision-level &
\multicolumn{1}{l}{\begin{tabular}[c]{@{}l@{}}

$\bullet$ Fusion of final outputs (e.g.,\\ probabilities) from unimodal models\\
$\bullet$ Tolerant to temporal asynchrony\\ and model heterogeneity\\
$\bullet$ Low computational cost for cross\\-modal communication

\end{tabular}} &
\multicolumn{1}{l}{\begin{tabular}[c]{@{}l@{}}

$\bullet$ Modalities with strong complementarity\\ and individual discriminative power\\
$\bullet$ Scenarios with asynchronous data or\\ different model architectures

\end{tabular}} &
\multicolumn{1}{l}{\begin{tabular}[c]{@{}l@{}}

$\bullet$ Loses fine-grained inter-modal interaction information\\
$\bullet$ Errors from early-stage processing are irreversible\\
$\bullet$ Performance is poor if individual modalities are weak

\end{tabular}}\\
\addlinespace[1em]

Hybrid &
\multicolumn{1}{l}{\begin{tabular}[c]{@{}l@{}}

$\bullet$ Dynamically integrates multiple\\ fusion levels, combining the\\ advantages of different strategies\\
$\bullet$ Hierarchical and progressive\\ fusion mechanism\\
$\bullet$ Retains specific details and\\ global consistency

\end{tabular}} &
\multicolumn{1}{l}{\begin{tabular}[c]{@{}l@{}}

$\bullet$ Complex multimodal tasks requiring a\\ combination of fusion advantages\\
$\bullet$ Scenarios needing flexible handling of\\ different modality characteristics\\
$\bullet$ Tasks with high demand for multi-level\\ information integration

\end{tabular}} &
\multicolumn{1}{l}{\begin{tabular}[c]{@{}l@{}}

$\bullet$ Requires balancing contributions from each fusion level\\
$\bullet$ Risk of information overload or computational redundancy\\
$\bullet$ Higher demands on model structure and training strategies

\end{tabular}}\\
\bottomrule
\end{tabular}}
\end{table*}

\subsubsection{Data-level fusion} \label{subsec:data_level}

\begin{figure*}[th!]
	\centering 
	\includegraphics[width= 1 \textwidth]{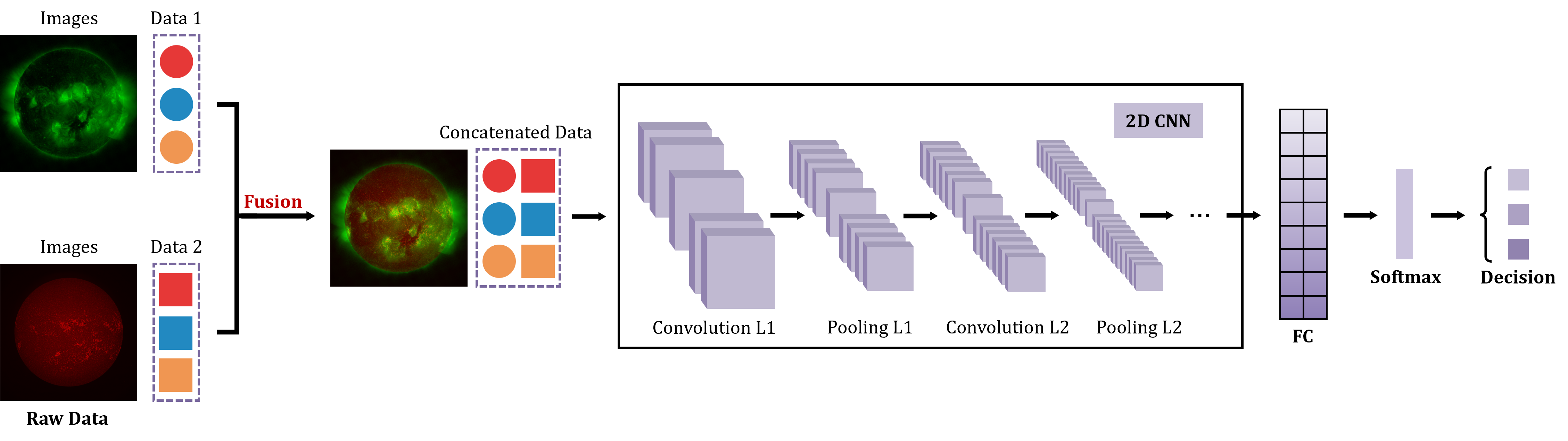}	
	\caption{Data-level fusion strategy. The raw data consists of solar observation image data in the extreme ultraviolet (green) and ultraviolet (red) bands.
    } 
	\label{fig:data-level}
\end{figure*}

Data-level fusion, also known as early fusion or pixel-level fusion \citep{shoeibi2023diagnosis, stahlschmidt2022multimodal, zhao2024deep, hong2020more, feng2020deep, ramachandram2017deep}, represents one of the earliest approaches to multimodal data integration \citep{baltruvsaitis2018multimodal}.  As the most fundamental fusion strategy, it directly integrates raw data from different modalities. This largely determines that it is suitable for cases where there is a high correlation between the modalities \citep{baltruvsaitis2018multimodal}, such as the integration of astronomical images captured across different wavebands. Figure \ref{fig:data-level} illustrates the schematic of this fusion strategy. The defining characteristic of data-level fusion lies in the fact that the integration occurs prior to model input, allowing subsequent feature extraction and model training to operate within a unified feature space. This design helps preserve the original information to the greatest extent possible and avoids the information loss that can arise from sequential processing \citep{zhao2024deep}. Furthermore, because the fusion is performed at the very beginning of the pipeline, the entire system requires training only a single model, thereby simplifying the development process \citep{steyaert2023multimodal}.

In practice, data-level fusion is typically implemented by concatenating or stacking multimodal raw data along a specific dimension (e.g., the channel dimension of images) to form high-dimensional feature vectors \citep{huang2020fusion}. Although this approach is simple and straightforward, it struggles to address the semantic and statistical heterogeneity across modalities, which limits its ability to exploit the complementary potential of multimodal information. Different modalities inherently vary in their information structures, and simple concatenation fails to achieve effective cross-modal interaction or alignment. Moreover, this strategy often assumes that all modalities are equally important and mutually independent, overlooking modality-specific weight differences and inter-modal dependencies that commonly arise in real-world applications \citep{steyaert2023multimodal}. Such oversimplification weakens the model's ability to learn joint representations and may introduce noise or redundancy in high-dimensional settings, even leading to the ``curse of dimensionality'' \citep{stahlschmidt2022multimodal, zhao2024deep}. Although several studies have attempted to mitigate these issues, the inherent constraints of it continue to pose challenges for effective multimodal integration. Therefore, this strategy is not currently accepted in astronomical MDF. In other disciplines, it is generally limited to highly customized tasks or combined with other fusion strategies. Nevertheless, there is still insufficient evidence to suggest that alternative fusion strategies consistently outperform data-level fusion in overall performance \citep{ramachandram2017deep, steyaert2023multimodal}.

\subsubsection{Feature-level fusion}

\begin{figure*}[th!]
	\centering 
	\includegraphics[width= 1 \textwidth]{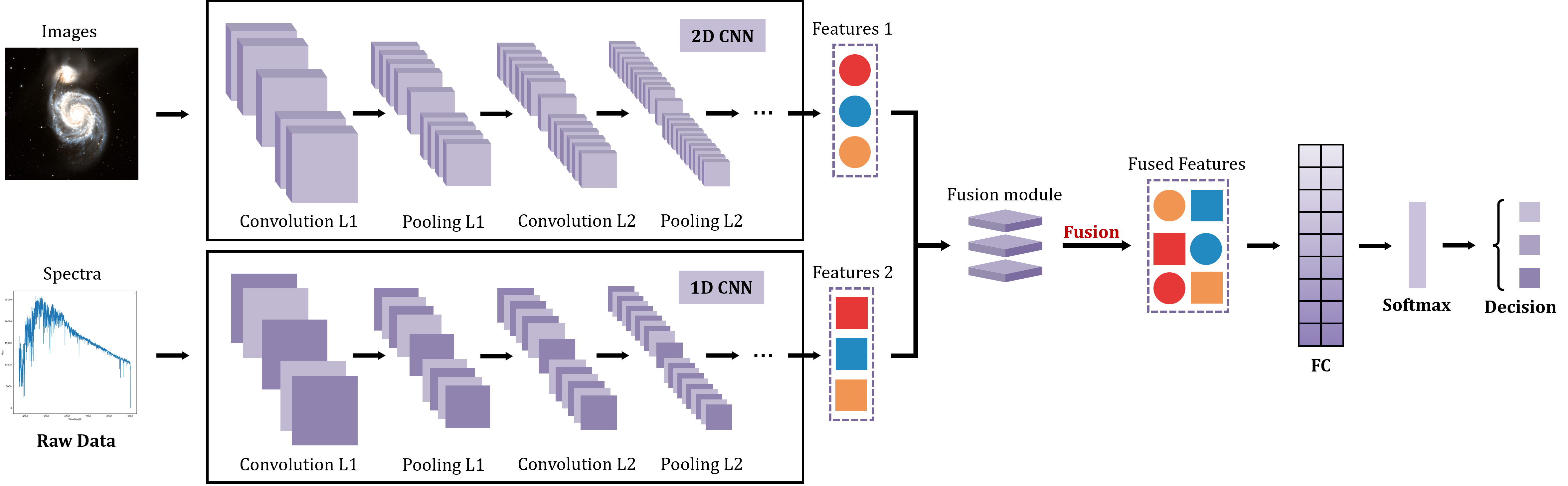}	
	\caption{Feature-level fusion strategy. The raw data consists of image data and spectral data of the M51 galaxy.} 
	\label{fig:feature-level}
\end{figure*}

Faced with the limitations of data-level fusion, a natural idea is to draw on the essence of DL, namely, extracting higher-level feature representations from raw data. This leads to the feature-level fusion strategy in MDF, also known as intermediate-layer fusion \citep{stahlschmidt2022multimodal, zhao2024deep, hong2020more, feng2020deep}. As shown in Figure \ref{fig:feature-level}, this strategy is typically implemented at the intermediate layer of a multimodal framework, positioned between the modality-specific feature extraction and the final decision-making phase \citep{gandhi2023multimodal, krones2025review}. In practice, each modality first passes through a dedicated subnetwork to extract modality-specific features. These features are then fed into a fusion module to a unified and more expressive joint feature representation, which is subsequently passed through FCs and a softmax function for task-specific prediction \citep{shoeibi2023diagnosis}. Unlike data-level fusion, which primarily focuses on the direct correlations among modalities, feature-level fusion emphasizes their complementary nature \citep{stahlschmidt2022multimodal, baltruvsaitis2018multimodal}.

In designing feature-level fusion methods, three key factors must be balanced: the degree of modality information preservation, the capacity to model inter-modal interactions, and computational cost. Based on different trade-off strategies, researchers employ various fusion methods. When the objective is to preserve as much original modality information as possible, concatenation serves as a common and effective choice. Pinciroli Vago \& Fraternali \citep{pinciroli2023deepgravilens} directly concatenated astronomical image features, time-series features, and statistical features extracted from time-series to construct multimodal combined features for gravitational lens classification, validating the effectiveness of this method in scenarios with well-aligned modalities and relatively simple structures. Feng et al. \citep{feng2025morpho} and Huijse et al. \citep{huijse2025learning} extended this method to higher-dimensional scenarios, obtaining more informative joint feature representations, thereby achieving superior performance in celestial source classification tasks. Nevertheless, concatenation remains a coarse-grained strategy, inherently assuming that modal information is parallel and non-competitive. Consequently, other intuitive fusion methods have been explored, including Discriminant Correlation Analysis (DCA) \citep{you2024applying}, weighted summation/averaging \citep{rizhko2025astrom3}, and element-wise operations \citep{ma2017multimodal, ethiraj2022classification, hosseinzadeh2024toward, cabrera2024atat, zhang2025white}. However, these methods still fail to account for the inherent nonlinear relationships between different modalities. To address this, Chen et al. \citep{chen2023radio} innovatively introduced a feedforward neural network to fuse text and image embeddings nonlinearly, allowing the model to automatically learn complex inter-modal dependencies without manually predefined combination rules. Yet, this method still lacks flexibility in weighting different modalities. Further advancing this direction, Junell et al. \citep{junell2025applying} proposed a gating-based multimodal framework, in which a gating network adaptively weights image and spectral features via learned weights, thereby dynamically modulating the contribution of each modality to the final classification decision. As research increasingly emphasizes the modeling of complex patterns and global dynamic weight adjustment, attention mechanism-based fusion has gained prominence. Zhao et al. \citep{zhao2025deep} incorporated the attention mechanism to dynamically capture intricate relationships among features from three modalities, substantially improving the accuracy of stellar spectral classification and demonstrating the power of attention in astronomical MDF. In practice, techniques such as contrastive learning \citep{zhang2024maven} and projection mechanisms \citep{wei2023identification, kamai2025machine} are often integrated to further enhance the performance of attention mechanism-driven fusion methods.

Undoubtedly, feature-level fusion has emerged as the core strategy for astronomical MDF, particularly suited for scenarios where modalities exhibit structural heterogeneity but strong semantic consistency \citep{mishra2024paperclip}. This strategy enables modality-specific feature extraction based on distinct physical properties and facilitates deep cross-modal interactions through efficient fusion mechanisms. By balancing performance and computational efficiency, it provides crucial support for large-scale surveys \citep{junell2025applying}. However, several limitations remain. First, fusion quality heavily relies on the representational capabilities of each modality's encoder. Insufficient feature extraction allows errors to propagate through the fusion module, making them difficult to correct. Second, feature-level fusion typically assumes a certain degree of semantic alignment among modalities. This assumption may fail when handling asynchronous observations, missing data, or significant spatio-temporal resolution differences, potentially yielding performance inferior to unimodal methods \citep{sun2022accurate, gao2023deep, siudek2025euclid}. Moreover, a growing number of researchers are expanding the number of modalities to three or even more \citep{ma2017multimodal, junell2025applying, rizhko2025astrom3}. Although the computational efficiency of feature-level fusion has been optimized compared to data-level fusion, careful architectural design is still required to prevent feature redundancy or dimensional collapse when combining high-dimensional embeddings.

\subsubsection{Decision-level fusion}

\begin{figure*}[th!]
	\centering 
	\includegraphics[width= 1 \textwidth]{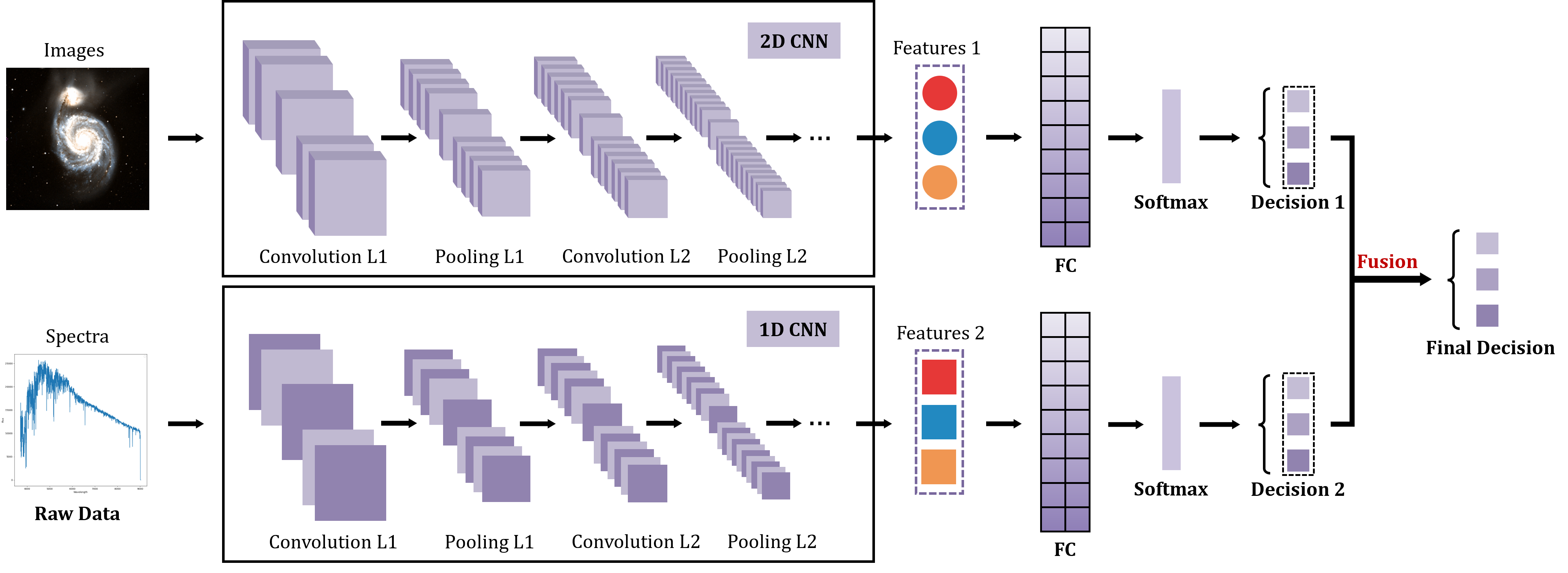}	
	\caption{Decision-level fusion strategy. The raw data consists of image data and spectral data of the M51 galaxy.} 
	\label{fig:decision-level}
\end{figure*}

Inspired by ensemble learning, decision-level fusion has also been proposed \citep{yang2025multimodal, ramachandram2017deep, tang2021multiple, hosseinzadeh2025end}. Decision-level fusion, also called late fusion \citep{stahlschmidt2022multimodal, zhao2024deep, hong2020more, feng2020deep}, is a high-level fusion strategy in MDF that combines the outputs (such as classification probabilities, regression predictions, or semantic labels) of independently processed modalities to generate a more robust global decision \citep{gandhi2023multimodal}. Figure \ref{fig:decision-level} illustrates the schematic diagram of decision-level fusion. Unlike data-level and feature-level fusion, decision-level fusion does not operate on raw data or intermediate features; instead, it allows each modality to complete a full inference process locally, achieving information fusion solely through backend predefined integration mechanism \citep{zhao2024deep}. The advantage of this strategy lies in its compatibility with the temporal asynchrony of data (e.g., differences in sampling times between spectral data and light curve data) and the heterogeneity of models (e.g., architectural differences between text models and image visual models) \citep{ramachandram2017deep}, while significantly reducing the computational overhead of cross-modal communication. 

Current implementations of decision-level fusion can be broadly categorized into two types: rule-driven methods (e.g., voting, weighted averaging) and model-driven methods (e.g., bagging, boosting, MLPs) \citep{yang2025multimodal, ramachandram2017deep, baltruvsaitis2018multimodal, zhao2024deep}. The former integrates the output results of different models through explicit rules, while the latter relies on machine learning models to automatically learn the optimal fusion method. In early studies, Hosseinzadeh et al. \citep{hosseinzadeh2024toward} proposed the Univariate Score Concatenation (USC) method, which concatenates the independent output scores of models based on proton flux time series and solar X-ray images, providing cross-modal decision evidence for the final classifier. Subsequently, Tang et al. \citep{tang2021multiple} employed the XGBoost algorithm to fuse the independent classification results of multiple neural networks, significantly improving the accuracy and robustness of magnetic-type classification for sunspot groups. Furthermore, in a later study, Hosseinzadeh et al. \citep{hosseinzadeh2025end} systematically evaluated various model-driven fusion approaches (e.g., Random Forests, Support Vector Machines, CatBoost, logistic regression, and MLPs) and analyzed their performance differences in predicting high-impact solar energetic particle events. These efforts not only verified the feasibility of decision-level fusion in complex astronomical tasks but also provided valuable insights for selecting appropriate fusion models in different application scenarios.

At first glance, decision-level fusion appears to offer clear advantages in both integration flexibility and implementation complexity, seemingly without limitations. However, this is not the case. Since the fusion operation occurs at the final stage of the processing pipeline, any critical information lost during the early-stage feature extraction or training of unimodal models will propagate directly to the final fused output, making it difficult to compensate for through subsequent fusion processes \citep{hosseinzadeh2024toward}. Worst of all, decision-level fusion cannot realize fine-grained interaction information between modalities; if the independent decision-making ability of each modality is weak, then the result of the fusion will not be ideal either \citep{yang2025multimodal, zhao2024deep, baltruvsaitis2018multimodal}. Thus, decision-level fusion is typically suitable for scenarios where modalities exhibit strong complementarity and each possesses strong independent discriminative capabilities \citep{hosseinzadeh2025end}. In applications requiring deeper cross-modal interactions, feature-level fusion often proves more effective \citep{ramachandram2017deep}.

\subsubsection{Hybrid fusion}

\begin{figure*}[th!]
	\centering 
	\includegraphics[width= 1 \textwidth]{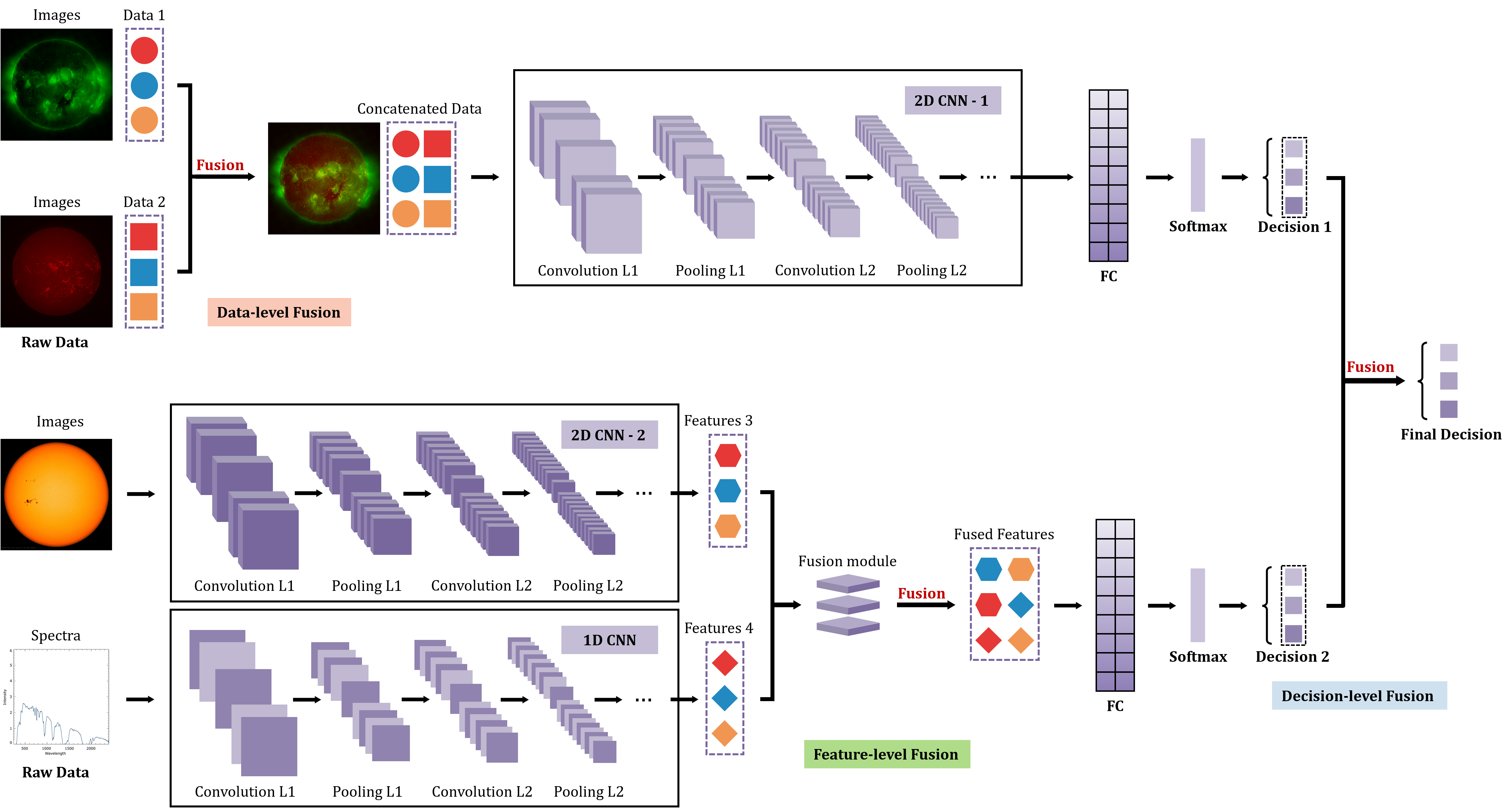}	
	\caption{Hybrid fusion strategy. The raw data in the data-level fusion consists of solar observation image data in the extreme ultraviolet (green) and ultraviolet (red). The raw data in the feature-level fusion consists of solar optical band image data and spectral data.} 
	\label{fig:hybrid}
\end{figure*}

Hybrid fusion is a comprehensive fusion strategy based on multi-level fusion design \citep{zhao2024deep, wang2023multi}. Figure \ref{fig:hybrid} illustrates the schematic diagram of hybrid fusion. Unlike being confined to a fixed level, hybrid fusion dynamically and flexibly integrates multiple fusion levels according to the characteristics of the task data and the requirements of the specific application \citep{baltruvsaitis2018multimodal}. This adaptability allows hybrid fusion to overcome the limitations of single-level approaches \citep{zhang2021deep}. Namely, it can leverage data-level fusion to preserve raw information, feature-level fusion to extract meaningful patterns, and decision-level fusion to refine the final output. This hierarchical and progressive fusion mechanism retains modality-specific details at lower levels while capturing global consistency at higher levels, making it a decent choice when addressing complex MDF task scenarios \citep{shaik2024survey, orchi2025contemporary}. For example, Junell et al. \citep{junell2025applying} implemented this concept in the AppleCiDEr framework by extracting features from four modalities (images, metadata, photometry, and spectra) through feature-level fusion, then dynamically integrating them via a Mixture-of-Experts (MoE) mechanism, and finally applying probabilistic averaging for decision-level fusion; similarly, Deng et al. \citep{deng2024ensemble} proposed the MESCR model, which fuses image and numerical features at the feature level and incorporates weighted sampling and ensemble learning at the decision level, significantly improving the accuracy of stellar classification and radius estimation. These works collectively demonstrate the effectiveness of multi-level coordinated fusion. It should be emphasized that when employing hybrid fusion strategy, it is necessary to balance the contributions of each fusion level to avoid information overload or computational redundancy caused by excessive levels \citep{zhang2021deep}. Besides, hybrid fusion requires stringent demands on the model structure and training strategies, necessitating end-to-end optimization or phased training to coordinate the parameters of different fusion modules, which places higher demands on model designers and training personnel.

\subsection{Model development} \label{subsec:development}

\begin{figure*}[th!]
	\centering 
	\includegraphics[width= 1 \textwidth]{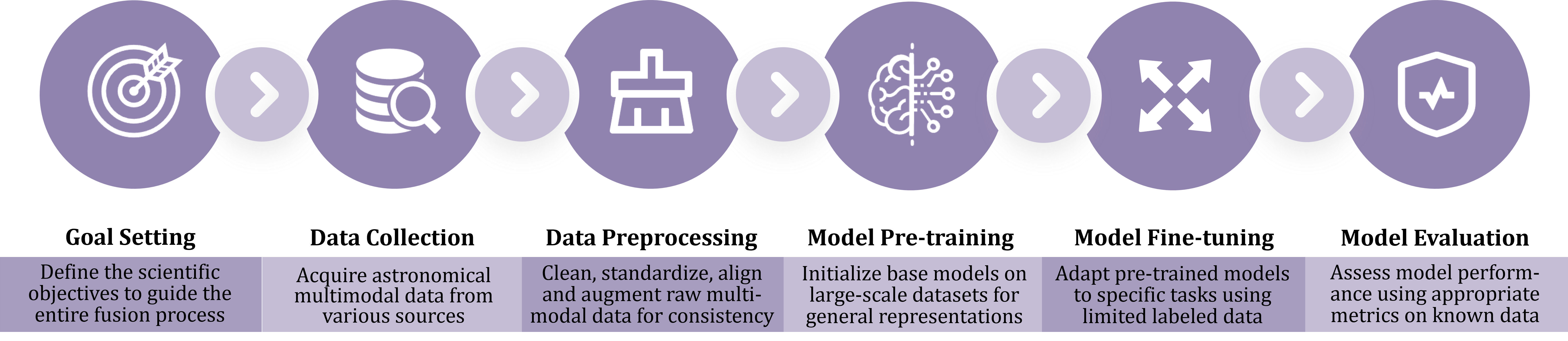}	
	\caption{The general pipeline of developing MDF models.
	} 
	\label{fig:pipeline}
\end{figure*}

The construction of multimodal learning models typically involves six phases: goal setting, data collection, data preprocessing, model pre-training, model fine-tuning, and model evaluation \citep{alegre2024identification}, as illustrated in Figure \ref{fig:pipeline}. Since goal setting varies significantly depending on specific tasks, it is difficult to describe uniformly. Data collection is also highly task-dependent, and relevant details have already been covered in Section \ref{sec:overview}, so it will not be repeated here. Next, we will only focus on the remaining four phases.

\subsubsection{Data preprocessing}

Data from different astronomical modalities vary in format, characteristics, and noise distribution. The goal of preprocessing is to transform these heterogeneous data into a consistent and model-processable form, laying the foundation for reliable fusion and analysis later on. Next, we will introduce the key steps of preprocessing, including data cleaning, normalization, alignment, and augmentation, and highlight their specific roles in astronomical MDF. Table ~\ref{tab:data_preprocessing} provides a summary.

\begin{table*}[ht!]
\centering
\caption{Summary of key preprocessing steps in astronomical MDF.}
\label{tab:data_preprocessing}
\footnotesize
\resizebox{\textwidth}{!}{
\begin{tabular}{@{}llll@{}}
\toprule
Step & Key objective & Common techniques & Role in MDF\\
\midrule

Data cleaning &
\multicolumn{1}{l}{\begin{tabular}[c]{@{}l@{}}Remove noise, outliers, and fill in\\ missing data \end{tabular}} &
\multicolumn{1}{l}{\begin{tabular}[c]{@{}l@{}}Isolation Forest; interpolation;\\ median filtering; Sigma Clipping\end{tabular}} &
\multicolumn{1}{l}{\begin{tabular}[c]{@{}l@{}}Improves data quality and reduces fusion bias \end{tabular}}\\
\addlinespace[1em]

Data normalization &
\multicolumn{1}{l}{\begin{tabular}[c]{@{}l@{}}Unify scales and numerical ranges \end{tabular}} &
\multicolumn{1}{l}{\begin{tabular}[c]{@{}l@{}}Min–max scaling, Z-score,\\ logarithmic transform\end{tabular}} &
\multicolumn{1}{l}{\begin{tabular}[c]{@{}l@{}}Ensures balanced feature contribution across\\ modalities \end{tabular}}\\
\addlinespace[1em]

Data alignment &
\multicolumn{1}{l}{\begin{tabular}[c]{@{}l@{}}Match data spatially, temporally,\\ and semantically \end{tabular}} &
\multicolumn{1}{l}{\begin{tabular}[c]{@{}l@{}}Coordinate cross-matching, image\\ registration, timestamp synchroni-\\zation, semantic annotation\end{tabular}} &
\multicolumn{1}{l}{\begin{tabular}[c]{@{}l@{}}Ensures spatial, temporal, and semantic\\ consistency for cross-modal association \end{tabular}}\\
\addlinespace[1em]

Data augmentation &
\multicolumn{1}{l}{\begin{tabular}[c]{@{}l@{}}Expand datasets and balance classes \end{tabular}} &
\multicolumn{1}{l}{\begin{tabular}[c]{@{}l@{}}Geometric transforms, noise\\ addition, generative models\end{tabular}} &
\multicolumn{1}{l}{\begin{tabular}[c]{@{}l@{}}Increases dataset size and diversity; improves\\ model generalization \end{tabular}}\\
\bottomrule
\end{tabular}}
\end{table*}

\textbf{Data cleaning}. Astronomical observations are often affected by factors such as instrumental noise, atmospheric disturbances, cosmic ray interference, and detector defects, introducing spurious signals and outliers into raw data. Data cleaning aims to restore the true celestial signals to the greatest extent possible, thereby enhancing the quality and reliability of multimodal data \citep{li2025deep, chen2023radio}. The common cleaning tasks include removing outliers (e.g., Isolation Forest \citep{siudek2025euclid}), filling in missing data (e.g., interpolation \citep{junell2025applying}, neighborhood reconstruction \citep{li2025estimation}), and suppressing instrumental and background noise (e.g., median filtering \citep{zhang2024maven}, Sigma Clipping \citep{alegre2024identification}). It is important to note that the cleaning process often requires integration with domain-specific knowledge, such as a precise understanding of instrument characteristics, observational conditions, and noise sources, to avoid introducing systematic biases during MDF.

\textbf{Data normalization}. Astronomical multimodal data often exhibit significant disparities in dimensionality and numerical ranges (e.g., image pixel values typically range from 0 to 255, while spectral fluxes may vary across multiple orders of magnitude \citep{liu2025addressing}). Without normalization, large-scale features dominate model loss functions and gradient updates, suppressing the learning of small-scale yet critical features. By mapping all modalities to a unified range (e.g., [0, 1] or [-1, 1]) through techniques such as min-max scaling \citep{sun2022accurate}, Z-score standardization \citep{junell2025applying}, or logarithmic transformation \citep{zhao2025finetuning, shen2025mixture}, feature magnitudes can be balanced. This ensures equitable learning across modalities, substantially improving the stability and convergence efficiency of multimodal training.

\textbf{Data alignment}. Astronomical multimodal data typically originate from observations under varying spatial, temporal, or instrumental conditions. To enable effective fusion of multimodal information, data alignment operations are required across three primary dimensions: spatial, temporal, and semantic. Spatial alignment relies on celestial coordinate information (e.g., right ascension and declination), employing techniques such as cross-matching \citep{li2019mcatcs} and image registration \citep{beroiz2020astroalign} to ensure different modalities correspond to the same observed target. Temporal alignment utilizes timestamp matching to synchronize data along the time axis, capturing cosmic phenomena at the same moment or within the same time interval \citep{zhao2025deep}. Semantic alignment aims to achieve content-level consistency between textual descriptions and other modalities, such as images or spectra. It serves as a crucial step for integrating textual data with observational data, and common approaches include manual annotation \citep{zaman2025astrollava}, citizen science projects \citep{chen2023radio}, and contrastive learning \citep{mishra2024paperclip}. It should be noted that the semantic alignment discussed here is performed at the raw data level, which differs from semantic alignment at the feature level in subsequent stages.

\textbf{Data augmentation}. The issues of data scarcity and imbalance can significantly constrain the generalization ability of astronomical multimodal models \citep{bowles2021attention}. Data augmentation techniques can expand training samples while maintaining astrophysical consistency \citep{wei2025photometric}. Currently, data augmentation is most commonly applied in image data processing, where methods such as cropping, rotation, flipping, or noise injection can simulate different observational conditions \citep{aniyan2017classifying, liu2023mfpim, alegre2024identification, parker2024astroclip}. For spectral or time-series data, perturbation processing can be adopted to enhance the model's resistance to interference \citep{cid2014resolving, hoyle2015data}. Deep generative models (e.g., VAEs, GANs, diffusion models) are also capable of simulating observational data of rare cosmic phenomena, providing further support for multimodal training \citep{zhao2023can, buck2024deep}.

\subsubsection{Model pre-training}

Model pre-training is often discussed with model fine-tuning, together forming a two-stage training paradigm that is widely adopted in the field of AI \citep{parker2024astroclip, zhang2024maven, walsh2024foundation}. Model pre-training typically refers to the initial training of a model on large-scale, weakly or unlabeled datasets through unsupervised or self-supervised learning tasks. This process employs specifically designed pre-training loss functions (e.g., contrastive or reconstruction losses) to guide the model in learning general-purpose data representations \citep{buck2024deep, rizhko2025astrom3, huijse2025learning}. The primary objective of this phase is not to directly optimize performance on specific downstream tasks but to enable the model's core components (including unimodal encoders and cross-modal fusion modules) to acquire two essential capabilities through parameter learning. The first is to extract high-quality, general, and robust feature representations for each modality. The second is to establish and strengthen semantic alignment and deep correlations across different modalities \citep{gao2023deep}. It is worth noting that the choice of fusion strategy determines the stage at which information interaction occurs and how it couples with the model architecture, thereby significantly influencing the complexity of model pre-training.

In data-level fusion, the fusion operation is performed prior to model pre-training. Consequently, the network architecture must be capable of processing high-dimensional, multi-channel, and complex fused data. In contrast, the pre-training process in decision-level fusion is relatively straightforward: each modality-specific model can be trained independently without considering inter-modal interactions \citep{zhang2025white}. Therefore, the design focus can be placed on optimizing intra-modal feature extraction to ensure strong discriminability and generalization capability. It is thus evident that neither data-level nor decision-level fusion involves learning cross-modal interaction capabilities. Moreover, in practical applications, the feature extraction processes in these two strategies are often not implemented separately but are integrated within a unified backbone network \citep{ji2023systematic, hosseinzadeh2025end}. In other words, they typically do not rely on pre-trained models but instead learn to extract effective feature representations from raw data through end-to-end training on downstream tasks, thereby determining the parameters of the feature encoders. Hence, in the subsequent discussion on model pre-training and fine-tuning, we will focus on the methods and designs related to feature-level fusion.

Unlike the two aforementioned strategies, feature-level fusion enables modality interaction at the intermediate representation stage of the model. Therefore, it must preserve modality-specific characteristics while promoting semantic compatibility across modalities. To achieve this, pre-training is often employed as an independent step at this strategy. The mainstream approach is to pre-train each modality separately, followed by cross-modal alignment. To leverage existing representations and reduce computational cost, many studies (e.g., \citep{parker2024astroclip}, \citep{cui2025pist}) have adopted approaches that freeze or fine-tune only parts of pre-trained encoders. However, such strategies limit the adaptability of the feature space to fusion tasks and may exacerbate modality imbalance, causing weaker modalities to be marginalized in the final decision-making process. To mitigate such asymmetry, some studies have introduced tighter cross-modal coupling mechanisms. Wei et al. \citep{wei2023identification} proposed a Bi-level attention mechanism that explicitly models cross-modal dependencies during feature extraction, while Hong et al. \citep{hong2023photoredshift} achieved modality alignment by mapping one modality into the ``standard'' feature space of another. Nevertheless, the former significantly increases model optimization complexity, whereas the latter implicitly assumes a master-slave relationship among modalities, potentially leading to the loss of complementary information. Building on these works, Zhao et al. \citep{zhao2025deep} further improved the design by employing the Mamba architecture to jointly process F10.7 index sequences and solar image. Their results demonstrated that a unified architecture combined with cross-attention mechanisms can achieve more balanced multimodal representations. While these studies relying on later-stage alignment mechanisms can indeed address the issues to some extent, they fail to enable joint optimization of the models during the pre-training stage. To address this, Hong et al. \citep{hong2023photoredshift} proposed a representative solution: achieving tight alignment of photometric-spectral data within a unified feature space through multimodal representation learning during the feature transformation stage, thereby enabling end-to-end joint optimization of photometric-spectral feature transformation model, and redshift prediction model.

\subsubsection{Model fine-tuning}

The essence of model fine-tuning lies in transforming the general representational capability of a pre-trained model into task-specific understanding and predictive ability through limited labeled data \citep{rizhko2025astrom3}. Within the framework of MDF, the fine-tuning process becomes more intricate, requiring careful decisions on which model components should be updated. Specifically, modal feature encoders can be selectively fine-tuned based on the similarity between the source and target domains \citep{walsh2024foundation}, while fusion modules typically need to be trained from scratch to capture new cross-modal relationships \citep{wei2023identification, gao2023deep}. Consequently, optimization at this phase focuses on two main objectives: model parameter adjustment and task-specific output mapping optimization. The latter is relatively uniform in its implementation, usually by introducing or optimizing the final layer of the model (e.g., a classifier or regression head), and thus is not the focus of our discussion.

In the field of astronomical MDF, fine-tuning strategies exhibit diverse characteristics. This reflects different perspectives and explorations regarding the key question of how to achieve effective knowledge transfer. Most current studies prioritize maximizing downstream task performance, making full fine-tuning the dominant strategy. Ma et al. \citep{ma2017multimodal} were among the first to introduce this strategy in a DL-based astronomical multimodal fusion network: they applied structured regularization terms to constrain network weights and used backpropagation to update feature encoders, FCs, and classifiers simultaneously, achieving end-to-end global fine-tuning. Subsequently, Wei et al. \citep{wei2025photometric} proposed a staged fine-tuning scheme, where individual modality subnetworks were optimized separately before joint fine-tuning with a lower learning rate. This yields a more stable global optimization process but inevitably reduces the transfer efficiency of the pre-trained model's general feature representations. To address this, Kamai et al. \citep{kamai2025machine} augmented full fine-tuning with task-specific prediction heads and dedicated loss functions, improving the models' flexibility for complex astronomical tasks. However, full fine-tuning usually performs well only when the pre-trained model is lightweight or the downstream task closely matches the pre-training data. As model scales continue to grow, Parameter-Efficient Fine-Tuning (PEFT) has emerged as a more attractive alternative. It updates only a small set of newly added or critical parameters while achieving performance comparable to full fine-tuning. In this context, approaches that freeze the backbone network and update only task-related layers have gained popularity, particularly for astronomical scenarios with limited or unevenly distributed modality data \citep{liu2023mfpim, mishra2024paperclip, walsh2024foundation}. Building on this trend, Zhao et al. \citep{zhao2025finetuning} pioneered the application of Low-Rank Adaptation (LoRA) to fine-tuning large-scale astronomical multimodal foundation models. By embedding low-rank matrix updates into the SpecCLIP \citep{zhao2025specclip} backbone, they achieved efficient adaptation and rapid transfer to new spectral datasets.

In practical applications, the choice of fine-tuning method should be made flexibly according to data scale, modality balance, and task complexity. It is important to emphasize that model fine-tuning is inherently an iterative optimization process. To approach optimal multimodal performance, training strategies and model parameters must be continuously refined in response to the specific requirements of the astronomical task at hand.

\subsubsection{Model evaluation}

Model evaluation is a comprehensive and systematic performance assessment of the constructed multimodal model to ensure its effectiveness, reliability, generalizability, and other characteristics, in practical astronomical tasks \citep{krones2025review}. First, it is essential to select suitable evaluation metrics that align with the specific objectives of the task, such as classification accuracy, precision, recall, F1-score for object identification \citep{ma2017multimodal, tang2021multiple, liu2023mfpim}, or mean squared error (MSE), mean absolute error (MAE) and $R^2$ for regression tasks like parameter estimation \citep{hong2023photoredshift, cuoco2021multimodal, sun2022accurate, buck2024deep}. Given the prevalence of data imbalance and noise in astronomical data, specialized metrics such as the AUC-ROC curve \citep{fang2019deep, francisco2024multimodal} and average precision \citep{gupta2024radiogalaxynet, zhang2024maven} are necessary. Moreover, evaluation should be conducted on representative and well-curated test datasets \citep{angeloudi2025multimodal, gupta2024radiogalaxynet, alegre2024identification}, ideally with diverse distributions that reflect real-world astronomical conditions. Cross-validation techniques \citep{tang2021multiple, hosseinzadeh2024toward, liu2023mfpim} and ablation studies \citep{pinciroli2023deepgravilens, wang2024effective} are often employed to analyze the contribution of each modality and evaluate the model's generalizability. Visualization techniques like t-SNE or UMAP can map high-dimensional features into a lower space, providing insights into the model's inner workings and optimization \citep{wang2024effective, buck2024deep, rizhko2025astrom3, wei2025photometric}, or demonstrating the complementarity of features from different modalities \citep{hosseinzadeh2025end}. It is also critical to compare the performance of the multimodal model against unimodal baselines and existing state-of-the-art methods to highlight the benefits of data fusion \citep{ouahmed2024multimodality, wei2023identification}. Finally, computational efficiency and cost \citep{cabrera2024atat, walsh2024foundation, liu2025addressing, li2025estimation}, though less frequently emphasized, are increasingly important in astronomical contexts, where the volume, velocity, and variety of data continue to grow rapidly. Together, these aspects form a robust framework for evaluating the practical utility and scientific value of DL-based multimodal fusion models in astronomy.

\subsection{Multimodal studies} \label{subsec:study}

In recent years, multimodal studies have sprung up in astronomy. In this subsection, we will comprehensively sort out the related studies before October 2025. Since there are as many as 58 such studies, and in order to make it easier for readers to browse and select what they are interested in, we present them in the form of tables (Table ~\ref{tab:studies1}, ~\ref{tab:studies2} and ~\ref{tab:studies3}). It is important to note that in the tables, the symbol ``---'' indicates that the aspect was not explicitly mentioned in the paper; the term ``multiple comparison results'' in the Result column signifies that the paper includes numerous detailed comparative analyses that cannot be exhaustively summarized here.

\begin{table*}[htbp]
\centering
\caption{Studies of MDF in astronomy (Part I).}
\label{tab:studies1}
\resizebox{\textwidth}{!}{
}
\end{table*}

\begin{table*}[htbp]
\centering
\caption{Studies of MDF in astronomy (Part II, continued from Table ~\ref{tab:studies1}). }
\label{tab:studies2}
\resizebox{\textwidth}{!}{
%
}
\end{table*}

\begin{table*}[htbp]
\centering
\caption{Studies of MDF in astronomy (Part III, continued from Table ~\ref{tab:studies2}). }
\label{tab:studies3}
\resizebox{\textwidth}{!}{
%
}\\
  \bottomrule
\end{tabular}}
\end{table*}

\subsection{Multimodal datasets} \label{subsec:dataset}

Against the backdrop of the current booming development of multimodal research, the large-scale integration of multimodal data has emerged as a core element for enhancing research efficiency and depth. In this subsection, we will describe 6 available multimodal datasets before October 2025. Table ~\ref{tab:datasets} provides an overview of them. For datasets not explicitly named by their authors, we will assign appropriate names based on their data sources, characteristics, and other relevant information to facilitate subsequent reference. For information not explicitly mentioned in the papers or dataset descriptions, we will use the symbol ``---'' to indicate its absence.

\begin{table*}[th!]
\centering
\caption{Multimodal datasets in astronomy. }
\label{tab:datasets}
\resizebox{\textwidth}{!}{
\begin{tabular}{@{}llllllllll@{}}
\toprule
  Dataset&
  Reference&
  Year &
  Research objective &
  Image data &
  Spectral data &
  Time-series data &
  Tabular data &
  Text data &
  Link\\ \midrule
  SMTD &
  \multicolumn{1}{l}{\begin{tabular}[c]{@{}l@{}}\citep{fang2019deep} \end{tabular}} &
  \multicolumn{1}{l}{\begin{tabular}[c]{@{}l@{}}2019 \end{tabular}} &
  \multicolumn{1}{l}{\begin{tabular}[c]{@{}l@{}}Sunspot magnetic\\ type recognition \end{tabular}} &
  \multicolumn{1}{l}{\begin{tabular}[c]{@{}l@{}}Continuum\\ images, magnet-\\ogram images \end{tabular}}& 
  \multicolumn{1}{l}{\begin{tabular}[c]{@{}l@{}}--- \end{tabular}}& 
  \multicolumn{1}{l}{\begin{tabular}[c]{@{}l@{}}--- \end{tabular}}& 
  \multicolumn{1}{l}{\begin{tabular}[c]{@{}l@{}}--- \end{tabular}}& 
  \multicolumn{1}{l}{\begin{tabular}[c]{@{}l@{}}--- \end{tabular}}& 
  \multicolumn{1}{l}{\begin{tabular}[c]{@{}l@{}}\href{https://tianchi.aliyun.com/dataset/74779}{https://tianchi.aliyun.com/dataset/74779} \end{tabular}}\\
  \addlinespace[1em]
  MPILD &
  \multicolumn{1}{l}{\begin{tabular}[c]{@{}l@{}}\citep{ji2023systematic} \end{tabular}} &
  \multicolumn{1}{l}{\begin{tabular}[c]{@{}l@{}}2023 \end{tabular}} &
  \multicolumn{1}{l}{\begin{tabular}[c]{@{}l@{}}Space weather\\ forecasting \end{tabular}} &
  \multicolumn{1}{l}{\begin{tabular}[c]{@{}l@{}}Binary mask\\ images \end{tabular}}& 
  \multicolumn{1}{l}{\begin{tabular}[c]{@{}l@{}}--- \end{tabular}}& 
  \multicolumn{1}{l}{\begin{tabular}[c]{@{}l@{}}--- \end{tabular}}& 
  \multicolumn{1}{l}{\begin{tabular}[c]{@{}l@{}}Time-series\\ metadata\\ from masks \end{tabular}}& 
  \multicolumn{1}{l}{\begin{tabular}[c]{@{}l@{}}--- \end{tabular}}& 
  \multicolumn{1}{l}{\begin{tabular}[c]{@{}l@{}}\href{https://doi.org/10.7910/DVN/BKP1RH}{https://doi.org/10.7910/DVN/BKP1RH} \end{tabular}}\\
  \addlinespace[1em]
  RGZ-MD &
  \multicolumn{1}{l}{\begin{tabular}[c]{@{}l@{}}\citep{chen2023radio} \end{tabular}} &
  \multicolumn{1}{l}{\begin{tabular}[c]{@{}l@{}}2023 \end{tabular}} &
  \multicolumn{1}{l}{\begin{tabular}[c]{@{}l@{}}Radio galaxy\\ classification \end{tabular}} &
  \multicolumn{1}{l}{\begin{tabular}[c]{@{}l@{}}Radio images,\\ infrared images \end{tabular}}& 
  \multicolumn{1}{l}{\begin{tabular}[c]{@{}l@{}}--- \end{tabular}}& 
  \multicolumn{1}{l}{\begin{tabular}[c]{@{}l@{}}--- \end{tabular}}& 
  \multicolumn{1}{l}{\begin{tabular}[c]{@{}l@{}}--- \end{tabular}}& 
  \multicolumn{1}{l}{\begin{tabular}[c]{@{}l@{}}Text discussions\\ on RadioTalk\\ forum \end{tabular}}& 
  \multicolumn{1}{l}{\begin{tabular}[c]{@{}l@{}}\href{https://zenodo.org/records/7988868}{https://zenodo.org/records/7988868} \end{tabular}}\\
  \addlinespace[1em]
  MultimodalUniverse &
  \multicolumn{1}{l}{\begin{tabular}[c]{@{}l@{}}\citep{angeloudi2025multimodal} \end{tabular}} &
  \multicolumn{1}{l}{\begin{tabular}[c]{@{}l@{}}2024 \end{tabular}} &
  \multicolumn{1}{l}{\begin{tabular}[c]{@{}l@{}}Addressing multi-\\ple research needs\\ in astronomy \end{tabular}} &
  \multicolumn{1}{l}{\begin{tabular}[c]{@{}l@{}}Multi-channel and\\ hyper-spectral ima-\\ges from JWST, etc.\end{tabular}}& 
  \multicolumn{1}{l}{\begin{tabular}[c]{@{}l@{}}Spectra from\\ Gaia, SDSS,\\ etc.  \end{tabular}}& 
  \multicolumn{1}{l}{\begin{tabular}[c]{@{}l@{}}Multivariate time-\\series from TESS,\\ PLAsTiCC, etc. \end{tabular}}& 
  \multicolumn{1}{l}{\begin{tabular}[c]{@{}l@{}}Catalogs from\\ Gaia, PROVA-\\BGS, etc. \end{tabular}}& 
  \multicolumn{1}{l}{\begin{tabular}[c]{@{}l@{}}--- \end{tabular}}& 
  \multicolumn{1}{l}{\begin{tabular}[c]{@{}l@{}}\href{https://huggingface.co/MultimodalUniverse}{https://huggingface.co/MultimodalUniverse} \end{tabular}}\\
  \addlinespace[1em]
  MSFPD &
  \multicolumn{1}{l}{\begin{tabular}[c]{@{}l@{}}\citep{mansouri2024multimodal} \end{tabular}} &
  \multicolumn{1}{l}{\begin{tabular}[c]{@{}l@{}}2024 \end{tabular}} &
  \multicolumn{1}{l}{\begin{tabular}[c]{@{}l@{}}Solar flare\\ prediction \end{tabular}} &
  \multicolumn{1}{l}{\begin{tabular}[c]{@{}l@{}}PILs Raster, Pola-\\rity Convex Hull\end{tabular}}& 
  \multicolumn{1}{l}{\begin{tabular}[c]{@{}l@{}}--- \end{tabular}}& 
  \multicolumn{1}{l}{\begin{tabular}[c]{@{}l@{}}Multivariate time-\\series of properties\\ related to PILs \end{tabular}}& 
  \multicolumn{1}{l}{\begin{tabular}[c]{@{}l@{}}Flare class data\\ (flare historical\\ data) \end{tabular}}& 
  \multicolumn{1}{l}{\begin{tabular}[c]{@{}l@{}}--- \end{tabular}}& 
  \multicolumn{1}{l}{\begin{tabular}[c]{@{}l@{}}\href{https://doi.org/10.7910/DVN/BKP1RH}{https://doi.org/10.7910/DVN/4A7ORF} \end{tabular}}\\
  \addlinespace[1em]
  RadioGalaxyNET &
  \multicolumn{1}{l}{\begin{tabular}[c]{@{}l@{}}\citep{gupta2024radiogalaxynet} \end{tabular}} &
  \multicolumn{1}{l}{\begin{tabular}[c]{@{}l@{}}2024 \end{tabular}} &
  \multicolumn{1}{l}{\begin{tabular}[c]{@{}l@{}}Radio galaxy\\ and infrared\\ host detection \end{tabular}} &
  \multicolumn{1}{l}{\begin{tabular}[c]{@{}l@{}}ASKAP radio\\ images, AllWISE\\ infrared images \end{tabular}}& 
  \multicolumn{1}{l}{\begin{tabular}[c]{@{}l@{}}--- \end{tabular}}& 
  \multicolumn{1}{l}{\begin{tabular}[c]{@{}l@{}}--- \end{tabular}}& 
  \multicolumn{1}{l}{\begin{tabular}[c]{@{}l@{}}--- \end{tabular}}& 
  \multicolumn{1}{l}{\begin{tabular}[c]{@{}l@{}}--- \end{tabular}}& 
  \multicolumn{1}{l}{\begin{tabular}[c]{@{}l@{}}\href{https://doi.org/10.25919/btk3-vx79}{https://doi.org/10.25919/btk3-vx79} \end{tabular}}\\
  \bottomrule
\end{tabular}}
\end{table*}

\textbf{SMTD Dataset}. Fang et al. \citep{fang2019deep} developed the sunspot magnetic type dataset (SMTD) for automatic classification of magnetic types in sunspot groups based on SDO/HMI observations. The dataset covers 720s SHARP (Space-Weather HARP) data from May 2010 to May 2017 and includes two modalities: magnetograms, capturing magnetic field distributions in active regions, and continuum images, depicting sunspot morphology in white light. All data are stored in FITS format and preprocessed through grayscale normalization, resizing to 160×80 pixels, and PNG conversion for CNN input. Each sample is manually labeled according to the Mount Wilson classification into three magnetic types: unipolar (Alpha), bipolar (Beta), and complex multipolar (Beta-x). The dataset contains 6,696 Alpha, 8,828 Beta, and 3,646 Beta-x samples for magnetograms, and 5,481 Alpha, 7,993 Beta, and 2,744 Beta-x samples for continuum images, exhibiting class imbalance. SMTD supports DL–based identification of sunspot magnetic types, facilitating improved forecasting of solar eruptive activity.

\textbf{MPILD Dataset}. Ji et al. \citep{ji2023systematic} presented the magnetic polarity inversion line dataset (MPILD), a large-scale resource designed to support space weather forecasting and analysis. The dataset integrates radial magnetic field (B\_r) and line-of-sight (B\_LoS) magnetograms from the Solar Dynamics Observatory’s Helioseismic and Magnetic Imager (SDO/HMI), offering a comprehensive view of solar active regions. It spans nearly the entire Solar Cycle 24 (May 2010-March 2019) and includes 4,090 HMI Active Region Patch (HARP) series. MPILD provides binary masks of polarity inversion lines (PILs), regions of polarity inversion (RoPI), and their convex hulls. It also includes time-series metadata derived from these masks, describing physical and morphological properties such as PIL size, RoPI area, masked unsigned flux, fractal dimension, eigenvalues, convexity, and Hu moments. All data are provided in standardized formats (binary masks in PNG and metadata in CSV) allowing direct mapping to original magnetograms and enabling multimodal analysis of the structural and temporal evolution of PILs in relation to solar eruptive activity.

\textbf{RGZ-MD Dataset}. Chen et al. \citep{chen2023radio} developed the Radio Galaxy Zoo Multimodal Dataset (RGZ-MD) based on the Radio Galaxy Zoo (RGZ) project. The dataset includes 10,643 radio subjects, each paired with coordinate-matched radio and infrared images, where infrared heatmaps are overlaid with radio contours. It also incorporates threaded text discussions from the RadioTalk forum, containing volunteer-provided tags and comments. Although sparse and brief, these texts provide valuable contextual information complementing the visual data. The authors pre-processed the text by concatenating comments, cleaning tags, and merging them into 11 unified categories such as ``artefact'', ``asymmetric'', and ``double''. In addition, embeddings were extracted using pre-trained models (BERT for text and ViT for images) yielding 768-dimensional representations for each modality. Publicly available on Zenodo, RGZ-MD enables multimodal learning for automated classification of radio galaxy morphologies by integrating both visual and textual features.

\textbf{MultimodalUniverse Dataset}. Angeloudi et al. \citep{angeloudi2025multimodal} introduced the MultimodalUniverse dataset, a large-scale collection of multimodal astronomical data designed to advance machine learning research. The dataset comprises hundreds of millions of observations ($\approx$100 TB) and includes multi-channel and hyperspectral images, spectra, multivariate time series, and extensive scientific metadata. Data are sourced from major ground- and space-based telescopes such as HSC, JWST, Gaia, SDSS, and DESI, offering comprehensive coverage of diverse astrophysical objects and phenomena. The dataset provides benchmark tasks and baseline DL models aligned with current astrophysical practices, serving as a foundation for performance evaluation and method comparison. All subsets are available in standardized Hugging Face formats, supporting both streaming and local access. Additionally, built-in tools for cross-modal matching enable flexible generation of custom multimodal pairings for various research applications.

\textbf{MSFPD Dataset}. Mansouri et al. \citep{mansouri2024multimodal} proposed the Multimodal Solar Flare Prediction Dataset (MSFPD), a comprehensive resource integrating multiple data modalities derived from SDO HARPs. The dataset includes raster images of polarity inversion lines (PILs), vector representations of polarity convex hulls, and multivariate time-series of PIL-related properties, all organized in a supervised learning format. Each sample is labeled with its corresponding flare class (FQ, C, M, or X) using historical SDO-GOES integrated flare records from May 2010 to January 2019. By combining raster, vector, and time-series modalities, MSFPD supports multimodal learning for improved solar flare prediction. Preliminary experiments demonstrate that this multimodal approach uncovers new spatiotemporal patterns and correlations, leading to enhanced predictive accuracy and better strategies for mitigating solar activity impacts.

\textbf{RadioGalaxyNET Dataset}. Gupta et al. \citep{gupta2024radiogalaxynet} constructed the RadioGalaxyNET dataset for automated detection and localization of extended radio galaxies and their infrared hosts. It comprises 2,800 three-channel images, including two radio channels from the ASKAP telescope and one infrared channel from the WISE W1 band (\SI{3.4}{\micro\meter}). The dataset contains 4,155 annotated galaxy instances, each labeled with class information, bounding boxes, pixel-level segmentation masks, and keypoints marking infrared host positions. All annotations follow the COCO format, enabling standardized evaluation across object detection methods. By integrating state-of-the-art ASKAP radio observations with corresponding infrared data and detailed annotations, RadioGalaxyNET provides a valuable benchmark for developing and assessing computer vision algorithms in radio astronomy.

\section{Discussion} \label{sec:discussion}

\subsection{Main findings}

In the process of reviewing the astronomical multimodal studies, we have distilled some findings. In this subsection, we will present some of the main findings with a view to providing valuable references and inspiration for future research.

The concept of ``multimodality'' is interpreted and applied flexibly in astronomical research. In general, multimodality was typically understood as the fusion of different types of multimedia data (e.g., image, audio, text) \citep{chen2020survey}. However, in the field of astronomy, to meet the evolving demands and adapt to development, the concept has been further extended. Unlike the categorization of multimedia data, ``modality'' in astronomy is a more fine-grained concept. Even within the same medium, different modalities can exist. For example, Gupta et al. \citep{gupta2024radiogalaxynet} and Ouahmed et al. \citep{ouahmed2024multimodality} treated images captured at different wavebands as distinct modalities, while Zhao et al. \citep{zhao2025finetuning} regarded spectral data from LAMOST, Gaia, and DESI as separate modalities. This understanding allows modalities to be categorized as either homogeneous (e.g., image-image) or heterogeneous (e.g., image-text) \citep{wu2023multimodal}. Moreover, the term ``multimodality'' is generally used from a definitional perspective and has a macroscopic characteristic. In practice, astronomers do not strictly adhere to the literal meaning of the word ``multi-'' in multimodality. Even the fusion of just two modalities can be referred to as multimodality rather than necessarily being called ``bimodality'' \citep{sun2022accurate, rehemtulla2024zwicky}.

\begin{figure*}[th!]
	\centering 
	\includegraphics[width= 0.7 \textwidth]{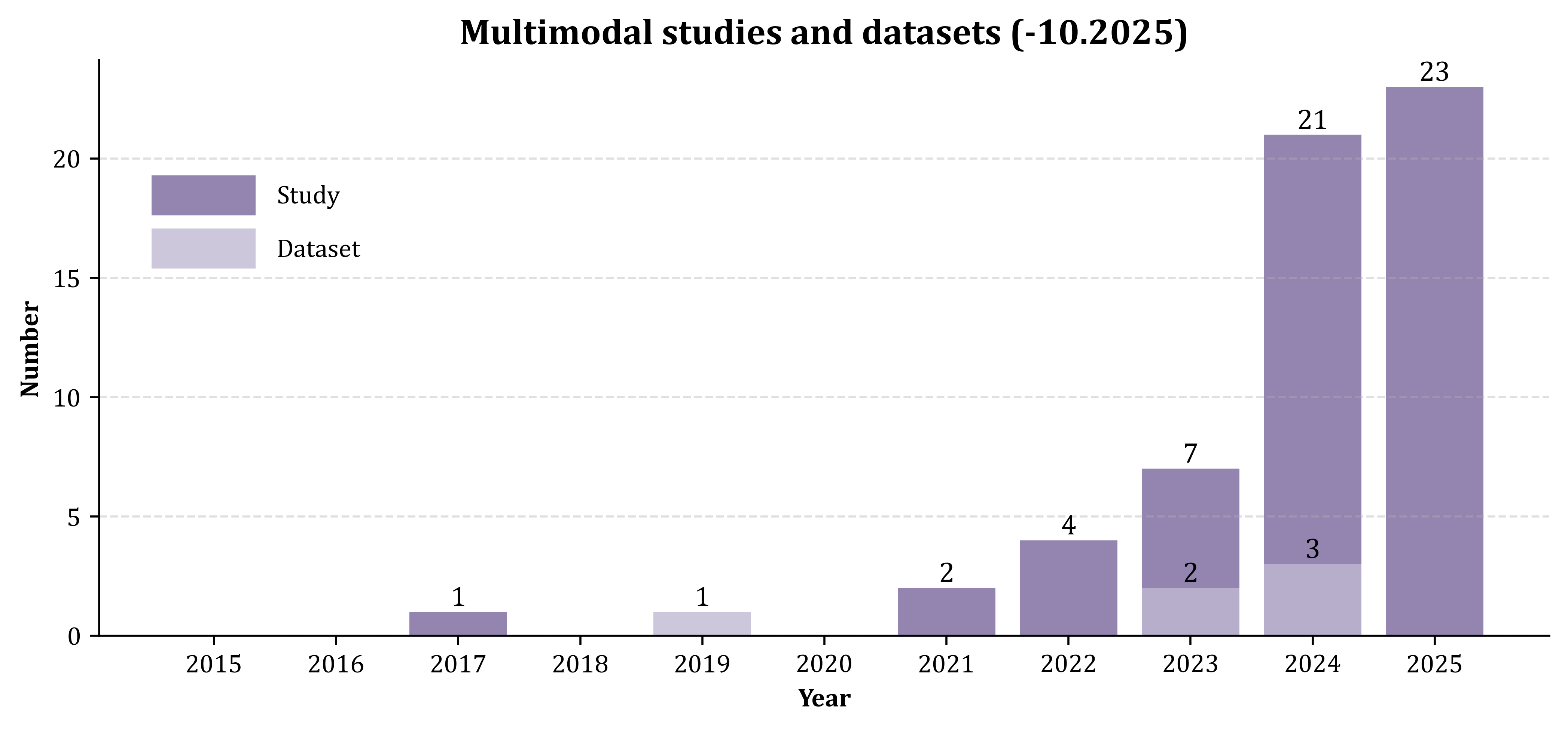}	
	\caption{ The number of studies and datasets on MDF in astronomy each year before October 2025.
	} 
	\label{fig:work}
\end{figure*}

Statistical analysis of the publication years in Tables ~\ref{tab:studies1}, ~\ref{tab:studies2}, and ~\ref{tab:datasets} clearly reveals a notable increase in the number of studies on astronomical multimodal research since 2023 (see Figure ~\ref{fig:work}). Prior to this, only a few pioneering works, represented by Ma et al. \citep{ma2017multimodal}, had explored this direction. We attribute this shift to breakthroughs in DL within computer vision (CV), NLP, and related areas. In particular, the emergence of (multimodal) large language models (LLMs) such as GPT-3.5, GPT-4, and Sora has not only popularized the concept of multimodality but also expedited the introduction and application of its underlying technologies in astronomy. At the same time, the systematic acquisition and timely open sharing of diverse observational data in astronomy have provided these models with rich experimental scenarios and validation foundations. At a deeper level, however, this trend is fundamentally driven by the astronomical community's strong demand for MDF \citep{junell2025applying}. The rise of multi-band and multi-messenger fusion paradigms embodies this intrinsic need \citep{yu2019astronomical}. Therefore, the rapid development of astronomical multimodal research can be viewed as the combined result of both external technological advancement and internal disciplinary evolution.

\begin{figure*}[th!]
	\centering 
	\includegraphics[width= 1 \textwidth]{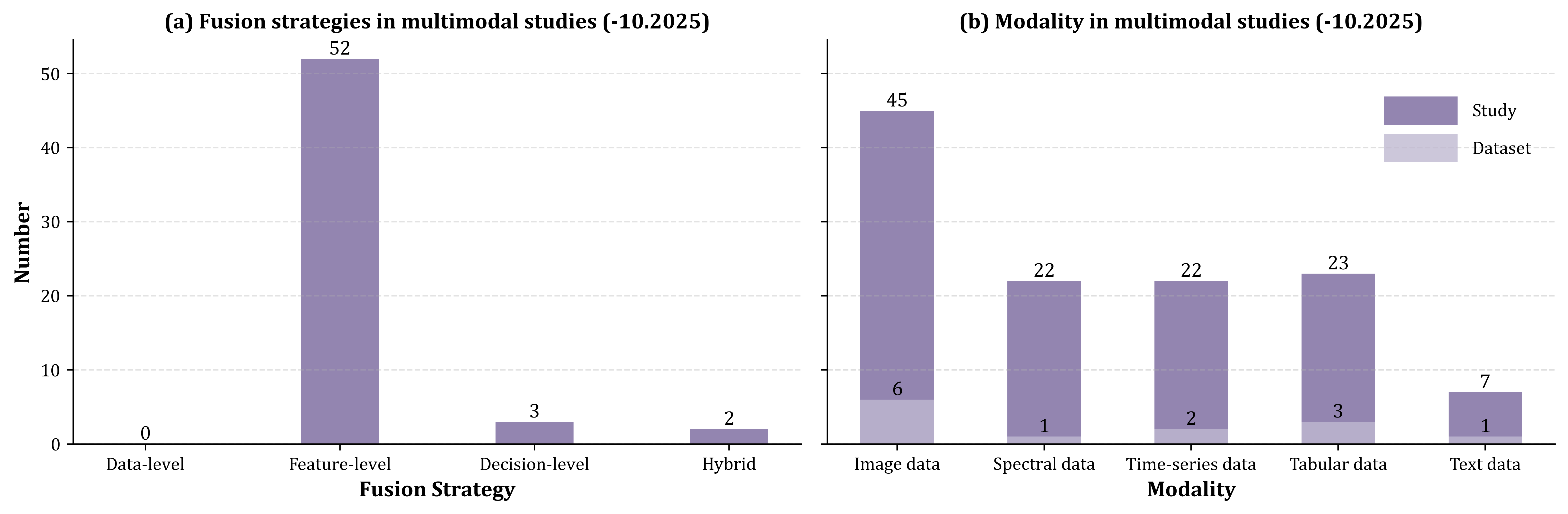}	
	\caption{(a): The selection of fusion strategies in studies of MDF in astronomy before October 2025; (b): The usage of modalities in studies and datasets of MDF in astronomy before October 2025.} 
	\label{fig:strategy_modality}
\end{figure*}

In summarizing the current state of multimodal research, we have identified several key differences. The vast majority of studies (over 93\%, 54/58, see Figure ~\ref{fig:strategy_modality} (a)) employ feature-level fusion as their core fusion strategy or partial fusion strategy (e.g., within hybrid fusion strategies). This dominance underscores the community’s growing emphasis on deep cross-modal interaction and, more broadly, demonstrates the strong potential of feature-level fusion in uncovering the intrinsic nature of complex cosmic phenomena. In contrast, decision-level fusion is exclusively applied in the field of space environment monitoring \citep{tang2021multiple, hosseinzadeh2024toward, hosseinzadeh2025end}. This reflects researchers' cautious attitude towards emerging technologies in the face of high real-time requirements, as well as their reliance on well-established unimodal techniques. Moreover, there is a significant imbalance in the use of various modalities in these studies. Image data dominate with an overwhelming proportion of approximately 78\% (45/58, see Figure~\ref{fig:strategy_modality} (b)). In comparison, spectral, time-series, tabular, and text data account for only about 38\%, 38\%, 40\%, and 12\%, respectively. This imbalance is similarly reflected in dataset construction. The current research status, which centers on visual data and marginalizes text data, is both a continuation of astronomy’s historical reliance on imaging data and an indication of the practical challenges in integrating unstructured text information.
However, recent advancements in cross-modal alignment techniques, such as the CLIP-based models utilized by Wang et al. \citep{wang2024effective} and Mishra-Sharma et al. \citep{mishra2024paperclip}, as well as the emergence of text-enhanced datasets like RGZ-MD \citep{chen2023radio}, suggest that text data are gradually gaining attention in astronomical MDF.

Solar physics exhibits notably leading research activity among all astronomical subfields, with 11 studies and 3 datasets involved, which can be evidenced from multiple perspectives. First, in terms of research volume, multimodal studies in solar physics have shown steady growth since the pioneering work of Ma et al. \citep{ma2017multimodal}. They cover a wide range of key topics, including sunspot group classification \citep{tang2021multiple}, solar flare forecasting \citep{francisco2024multimodal, li2024solar}, and solar flux prediction \citep{zhao2025deep}. During this process, the field has systematically developed several specialized datasets, including SMTD \citep{fang2019deep}, MPILD \citep{ji2023systematic}, and MSFPD \citep{mansouri2024multimodal}. Second, regarding fusion strategies, solar physics research not only widely adopts feature-level fusion but also actively explores decision-level fusion, particularly for tasks requiring high timeliness. Furthermore, this field benefits from a richer variety of data modalities, encompassing images (e.g., magnetograms \citep{li2024solar}), spectra (e.g., EVE ultraviolet spectra \citep{walsh2024foundation}), time series (e.g., proton flux \citep{hosseinzadeh2024toward}, F10.7 index \citep{zhao2025deep}), and tabular data (e.g., NOAA sunspot parameters \citep{wang2024solar}). This diversity enables more combinations of modalities and makes solar physics naturally more suitable for developing multimodal models than other subfields, which often rely on more homogeneous data types. It is also worth noting that the high level of research activity may be driven by urgent demands from major application scenarios such as space weather warning. Overall, the pioneering experience accumulated in solar physics provide valuable references for advancing MDF in other subfields.

\subsection{Challenges and prospects}

\textbf{Heterogeneous modality consistency and fusion}. The heterogeneity among astronomical data modalities poses a fundamental challenge to constructing a unified and reliable multimodal framework. While homogeneous modality data obtained within a single facility typically exhibit high consistency and can be readily fused after relatively simple preprocessing, real-world applications often involve complex heterogeneous modalities capturing multidimensional information. These data modalities often originate from instruments employing different physical mechanisms, sampling schemes, noise statistics, and coordinate systems, leading to systematic issues such as mismatched point spread functions (PSFs) and resolutions, as well as inconsistent temporal sampling rates. Although some preprocessing techniques can mitigate these discrepancies to some extent, the effects are often limited. This has forced many studies to assume that heterogeneous modality inputs can be directly concatenated or embedded into a shared latent space, ignoring the incompatibility of their underlying physical information. Only a few studies have focused on this issue. Recently, Jia et al. \citep{jia2025rapid} developed a multi-channel image calibration and fusion framework incorporating PSF data. Going forward, it will be crucial to embed specialized modality heterogeneity mitigation modules, such as cross-modal calibration, uncertainty-aware normalization, and physics-based resolution matching, directly into fusion architectures.

\textbf{Multimodal fusion benchmarks}. Large-scale, standardized, and systematically curated astronomical multimodal datasets are a fundamental prerequisite for developing powerful fusion foundation models. Although the astronomical community has collaboratively built extensive survey archives, research on MDF remains highly fragmented. Most studies are forced to construct small, task-specific datasets tailored to their own scientific goals. As a result, models optimized on narrowly defined datasets often fail to generalize to other scenarios. In contrast to the benchmark role of ImageNet in CV, astronomy has long lacked a unified benchmark to systematically drive algorithmic progress. To fill this gap, the MultimodalUniverse dataset, encompassing 100 TB of data, has emerged \citep{angeloudi2025multimodal}. This effort represents a shift from isolated, task-specific data silos toward a community-driven data platform. However, this is only a first step. Future progress requires coordinated community-wide collaboration to build more standardized, cross-survey benchmark datasets. Such benchmarks must adopt unified calibration procedures, consistent annotation protocols, and broad modality coverage to support the next generation of multimodal astronomical research. In this process, organizations such as the International Virtual Observatory Alliance (IVOA), astronomical data centers, and computational centers must take a leading role.

\textbf{Computational efficiency and scalability}. The increasing demands for MDF in current astronomical research have led to substantial storage and computational costs when processing large and complex multimodal datasets. The introduction of modules such as cross-modal alignment, attention-based fusion, and uncertainty modeling further exacerbates the computational burden, with requirements potentially growing exponentially or even worse with data scales. To operate within limited computational resources, many fusion studies often resort to heavy downsampling, truncated spectra, or simplified models, which lead to the loss of fine-grained physical information that should be retained. This contradicts the original intent of MDF. Li et al. \citep{li2025estimation} proposed developing modular fusion architectures that support distributed execution to mitigate these challenges, which is feasible for appropriately sized datasets. However, when constructing multimodal foundation models that require vast amounts of data in the future, global issues such as I/O bottlenecks and seamless integration with the astronomical data ecosystem will arise. The emerging ``cloud-first'' data access paradigm (e.g., NADC Cloud Platform \footnote{https://science.china-vo.org/sp/portal}, NOIRLab Astro Data Lab Science Platform \footnote{https://datalab.noirlab.edu}, ESA Data Lab \footnote{https://datalabs.esa.int}) offers a feasible path for ``data and computation without local landing''. In the future, cloud-native fusion algorithms should be systematically designed to directly interface with cloud storage, leverage serverless function elasticity, and rely on GPU-accelerated environments, transforming MDF from customized analysis to scalable, on-demand services for the entire astronomical community.

\textbf{Model interpretability}. Under the demands of scientific rigor, it is not only necessary to comprehend the predictions made by models but also to elucidate how specific cross-modal interactions drive the generation of these predictions. Current mainstream multimodal architectures, typically constructed upon deep feature encoders, cross-attention mechanisms, and high-dimensional latent representations, often obscure the underlying physical relationships between modalities. This makes it difficult to determine whether models rely on genuine astrophysical correlations or spurious shortcuts introduced by data artifacts or noise, a phenomenon clearly revealed by Huijse et al. \citep{huijse2025learning}. This issue is particularly pronounced when integrating modalities governed by distinct physical processes, directly affecting the reliability of cross-modal physical inferences. Despite its clear importance, interpretability remains underexplored in existing work. When discussed at all, it is usually treated superficially or relegated to a minor limitation in the discussion section, without detailed examination of feature-level mechanisms \citep{junell2025applying}. Moving forward, the field must shift towards a fusion paradigm guided by Explainable Artificial Intelligence (XAI). This includes introducing attention mechanisms constrained by physical information, layer-wise relevance propagation methods, modality-aware causal reasoning frameworks, attribution maps capable of decomposing uncertainties, and symbolic or generative surrogate models that enable scientific interrogation of latent representations.

\textbf{Data scarcity}. Data scarcity is a pressing challenge that needs to be addressed in both unimodal and multimodal contexts. Many cosmic phenomena of significant scientific value, such as fast transients, strong gravitational lensing systems, high-redshift galaxies, and multi-messenger events, occur so rarely that they cannot support data-hungry DL models \citep{hosseinzadeh2024toward}. Despite the vast amounts of raw observational data provided by survey projects, effective overlap between different modalities is extremely limited: high-resolution imaging regions often lack spectroscopic coverage, radio interferometric data are difficult to synchronize with optical time-domain surveys, and multi-messenger detections confirm only a handful of cross-modal correlation events each year. Consequently, most studies are forced to train models on narrow or incomplete multimodal samples, assuming that limited data can still represent the full diversity of astrophysical sources. This assumption restricts the models' generalization capabilities. Very recently, Shen et al. \citep{shen2025mixture} have attempted to mitigate this issue by using self-supervised pre-training on unimodal data and synthesizing missing modalities with generative models, but these approaches have not fundamentally solved the problem of learning with extremely low sample sizes. Following the practice of Scognamiglio et al. \citep{scognamiglio2025denoising}, systematically embedding physical priors into learning architectures such as diffusion models and developing cross-modal completion frameworks that accommodate observational uncertainties may hold promise.

\textbf{Open science and collaboration}. Compared to disciplines such as healthcare \citep{krones2025review}, the application of cutting-edge MDF technology in astronomy exhibits a more pronounced lag. This situation not only highlights the importance of open science but also emphasizes the urgency of interdisciplinary collaboration between astronomy and fields such as computer science and data science. To advance this process, we should strive to strengthen the practice of open science, which includes, but is not limited to, sharing datasets, models, and code resources. Creating open-source tools for MDF can lower the barrier to entry for new researchers and encourage the adoption of best practices across the field. Additionally, attempting to regularly organize challenge tasks related to astronomical MDF on open platforms like Kaggle \footnote{https://www.kaggle.com/}, similar to the ImageNet competition, can attract researchers from diverse disciplinary backgrounds to gather and jointly tackle astronomical scientific problems. This will greatly enrich the reserve force in this field, stimulate innovative vitality, and jointly promote the progress of MDF research in astronomy.

\section{Conclusion} \label{sec:conclusion}

The explosive growth of multimodal data in astronomy presents substantial challenges while also creating unprecedented opportunities for scientific discovery. By jointly exploiting the complementary physical information embedded in image, spectral, time-series, tabular, and text data resources, the field has made remarkable progress in recent years. The rise of DL, driven by its powerful representation learning capabilities and its ability to model cross-modal interactions, has further highlighted clear advantages over traditional analytical frameworks. These advancements have established DL-based MDF as one of the most active research directions in modern astronomy. This review provides a comprehensive review of this rapidly evolving area, covering data modality types, representative DL architectures, fusion strategies, model development practices, exemplary studies, public datasets, and emerging research trends. Particular emphasis is placed on DL-based fusion strategies and their diverse implementations across astronomical tasks. Encouragingly, multimodal fusion techniques are now integrated into many subfields of astronomy, achieving strong performance in source classification, parameter estimation, transient detection, and multi-messenger analysis. These developments collectively demonstrate the transformative potential of MDF and reinforce confidence within the community. Supported by open data initiatives, continued technical innovation, and interdisciplinary collaboration, future research will further advance our understanding of the universe.

\section*{CRediT authorship contribution statement}

\textbf{Wujun Shao:} Writing – review \& editing, Writing – original draft, Methodology, Visualization, Investigation. \textbf{Dongwei Fan:} Writing – review \& editing, Supervision, Investigation. \textbf{Chenzhou Cui:} Writing – review \& editing, Supervision, Investigation. \textbf{Yunfei Xu:} Writing – review \& editing, Investigation. \textbf{Shirui Wei:} Writing – review \& editing. \textbf{Xin Lyu:} Writing – review \& editing. 

\section*{Declaration of competing interest}

The authors declare that they have no known competing financial interests or personal relationships that could have appeared to influence the work reported in this paper.

\section*{Data availability}

No data was used for the research described in the article.

\section*{Acknowledgements}

This work is supported by the National Key R\&D Program of China (2022YFF0711500), National Natural Science Foundation of China (NSFC) (12273077, 12403102, 12373110, 12103070, and 1240030222), Strategic Priority Research Program of the Chinese Academy of Sciences (XDB0550101). Data resources are supported by China National Astronomical Data Center (NADC), CAS Astronomical Data Center and Chinese Virtual Observatory (China-VO). This work is supported by Astronomical Big Data Joint Research Center, co-founded by National Astronomical Observatories, Chinese Academy of Sciences and Alibaba Cloud.

\appendix
\section{Abbreviation reference table} \label{app:abbreviation}

This appendix provides a comprehensive reference table listing all abbreviations and full names of astronomical telescopes, sky surveys and observatories mentioned in this paper.

\begin{table*}[th!]
    \centering
    \caption{Abbreviations and full names of astronomical telescopes, sky surveys and observatories (sorted by abbreviation A--Z, row--first).}
    \label{tab:telescopes_surveys}
    \footnotesize
    \renewcommand{\arraystretch}{1.1}
    \resizebox{\textwidth}{!}{
    \begin{tabular}{@{}llllll@{}}
        \hline
        Abbreviation & Full name & & & Abbreviation & Full name \\
        \hline
        APOGEE & Apache Point Observatory Galactic Evolution Experiment & & & ASKAP & Australian Square Kilometre Array Pathfinder \\
        Auger & Pierre Auger Observatory & & & CFHTLS & Canada-France-Hawaii Telescope Legacy Survey \\
        Chandra & Chandra X-ray Observatory & & & DESI & Dark Energy Spectroscopic Instrument \\
        EIT & Extreme-ultraviolet Imaging Telescope & & & EP & Einstein Probe \\
        Euclid & Euclid Space Telescope & & & FAST & Five-hundred-meter Aperture Spherical Radio Telescope \\
        Fermi LAT & Fermi Gamma-ray Space Telescope Large Area Telescope & & & Gaia & Global Astrometric Interferometer for Astrophysics \\
        GOES & Geostationary Operational Environmental Satellites & & & GAMA & Galaxy And Mass Assembly \\
        HESS & High Energy Stereoscopic System & & & HSC & Hyper Suprime-Cam \\
        HST & Hubble Space Telescope & & & IceCube & IceCube Neutrino Observatory \\
        JWST & James Webb Space Telescope & & & K2 & Kepler/K2 Mission \\
        Keck & W. M. Keck Observatory & & & Kepler & Kepler Space Telescope \\
        KiDS & Kilo-Degree Survey & & & LAMOST & Large Sky Area Multi-Object Fiber Spectroscopic Telescope \\
        LHAASO & Large High Altitude Air Shower Observatory & & & LIGO & Laser Interferometric Gravitational-wave Observatory \\
        LOFAR & Low Frequency Array & & & Pan-STARRS & Panoramic Survey Telescope and Rapid Response System \\
        Parker & Parker Solar Probe & & & SDO & Solar Dynamics Observatory \\
        SDSS & Sloan Digital Sky Survey & & & SKA & Square Kilometre Array \\
        SOHO & Solar and Heliospheric Observatory & & & Swift & Neil Gehrels Swift Observatory \\
        TESS & Transiting Exoplanet Survey Satellite & & & VLT & Very Large Telescope \\
        Virgo & Virgo Gravitational Wave Interferometer & & & VRO & Vera C. Rubin Observatory \\
        WISE & Wide-field Infrared Survey Explorer & & & ZTF & Zwicky Transient Facility \\
        \hline
    \end{tabular}}
\end{table*}

\bibliographystyle{elsarticle-harv}
\bibliography{example}

\end{document}